\documentclass[twocolumn, showpacs, preprintnumbers, amsmath,amssymb]{revtex4}
\usepackage[colorlinks,urlcolor=blue]{hyperref}
\usepackage{graphicx}
\usepackage{lineno}
\usepackage{dcolumn}
\usepackage{bm}
\usepackage{appendix}
 \usepackage{verbatim}
\newcommand{\reform}{\color{black} }

\newtheorem{remark}{Remark}

\newcommand{\sign}{\mathrm{sign}\,}
\usepackage{breakurl}
\usepackage{amsmath}
\usepackage{enumerate}
\usepackage[draft]{changes}     
\definechangesauthor[name={Author}, color=red]{Author}

\makeatletter
  \let\@font@info\@gobble
  \let\@font@warning\@gobble
\makeatother

\begin{document}
\preprint{APS/123-QED}

\title{Renewal theory with fat tailed distributed sojourn times: typical versus rare}

\author{Wanli Wang$^{1,2}$}

\author{Johannes H. P. Schulz$^2$}

\author{Weihua Deng$^{1}$}%

\author{Eli Barkai$^{2}$}%
\affiliation{%
  $^1$School of Mathematics and Statistics, Gansu Key Laboratory of Applied Mathematics and Complex
Systems, Lanzhou University, Lanzhou 730000,  P.R. China\\
$^2$Department of Physics, Institute of Nanotechnology and Advanced Materials, Bar-Ilan University, Ramat-Gan
52900, Israel
\\
 }%


\date{\today}

\begin{abstract}
 Renewal processes with heavy-tailed power law distributed sojourn times
are commonly encountered in physical modelling
and so typical fluctuations of observables
of interest have been investigated in detail. To describe rare events
the rate function approach from large deviation theory does not hold and
new tools must be considered. Here we investigate the large deviations
of the number of renewals, the forward and backward recurrence time,
the occupation time, and the time interval straddling  the observation time.
We show how non-normalized densities describe these rare fluctuations,
and how moments of certain observables are obtained from these limiting
laws. Numerical simulations illustrate our results showing the deviations from arcsine, Dynkin, Darling-Kac, L{\'e}vy and Lamperti  laws.
%
%
%



\end{abstract}

\pacs{02. 50. -r,  05. 20. -y,  05. 40. -a }
%

%


\maketitle

\section{Introduction}\label{sect1}

 Renewal processes \cite{Godreche2001Statistics,Mainardi2004fractional,Mainardi2007Beyond,Godreche2015Statistics,Niemann2016Renewal,Akimoto2016Langevin} are simple stochastic models   for events that occur on the time axis when the time intervals between events are independent and identically
distributed (IID) random variables. This idealised approach has many applications,
ranging from the analysis of photon arrival times to queuing  theory.
In some models the sojourn time probability density function (PDF)
has fat tails, and this leads to fractal time renewal processes.
In the case when the variance of the sojourn time diverges, we have
deviations from the normal central limit theorem
and/or the law of large numbers. Such fat tailed processes are observed
in many systems, ranging from blinking quantum dots \cite{Simone2005Fluorescence}, to diffusion
of particles in polymer  networks \cite{Edery2018Surfactant},
or diffusion
of particles on the membrane of cell \cite{weron2017Ergodicity} to name a few.
In these systems
the renewal process is triggering jumps in intensity or in space.
The continuous time random walk model \cite{Metzler2000random}, the annealed trap model,
the zero  crossing  of Brownian motion, the velocity zero crossing
of cold atoms diffusing in momentum space \cite{Barkai2014From}, are all well known
models which use this popular renewal approach
(see however \cite{Boettcher2018Aging,Nyber2018Zero}).
Heavy tailed renewal theory is also used in the context of  localization   in random waveguides \cite{Fernandez2014Beyond}.
The number of renewals, under certain conditions, is described
by L\'evy statistics, and the fluctuations in these
processes are large. Hence it is important to explore the
rare events or the far tails of the distributions of
observables of interest.
As
mentioned  in \cite{Touchette2009large,Whitelam2018Large} the large deviation principle, with its characteristic exponential decay of large fluctuations, does not describe this case, and
instead the big jump principle \cite{Alessandro2018Single} is used to evaluate the rare events in L{\'e}vy type of processes.

The main statistical tool describing observables
of interest are non-normalised states, which are limiting laws
with which we may obtain statistical information on the system, 
including for example the variance, which in usual circumstances is the way we measure fluctuations.
These non-normalised states were previously investigated, in the
context of L\'evy walks \cite{Rebenshtok2014Infinite},
 spatial diffusion of cold atoms \cite{Erez2017Large},
 and very recently  for Boltzmann-Gibbs states
 when the underlying partition
function of the system  diverges \cite{Erez2018From}. These
functions describing the statistical behavior of the
system  are sometimes
called infinite densities or infinite covariant densities, and
they appear constantly in infinite ergodic theory \cite{Aaronson1997introduction}.

   Our goal in this paper is to investigate the statistics of rare
events in renewal theory. Consider for example a non-biased ordinary
random walk on the integers. The spatial jump
process is Markovian hence the zero
crossing, where the zero is the origin,  is a renewal process.
Here like Brownian motion, the waiting time PDF between the zero crossings is fat tailed,
in such a way that the mean return time diverges.
The distribution of the  occupation time $0<T^{+}<t$,
namely the time  the random walker
spends in the  positive domain is well investigated \cite{Redner2001guide,Godreche2001Statistics}.
Naively one would expect that when the  measurement
time $t$ is long the particle will spend half of its time
to the right of the origin. Instead one finds that this is the least likely
scenario, and the  PDF  of the properly scaled occupation time reads
\begin{equation}
\lim_{t\to\infty}f_{T^{+}/t} (x) = {1 \over \pi\sqrt{ x (1 - x)}}.
\label{eq0011}
\end{equation}
Here and all along this manuscript the subscript denotes
the observable of interest, e.g.  we consider the PDF of $T^{+}/t$ which
attains values $0<x<1$.
This arcsine law, which describes also other features of Brownian motion \cite{Morters2010Brownian,Akimoto2016Distributional,Sadhu2018Generalized},
exhibits divergences on $x\to 0$ or $x\to 1$.
Here a particular scaling of $T^{+} \propto
 t$ is considered. However, in cases studied
below we show that other limiting laws are
found when a second time scale is  considered
and these may modify the statistical properties of the occupation
time  when
$T^{+}$ is either very small or very large.
This in turn influences
the anticipated blow up of the arcsine law at its extremes. Notice that here the
least likely event, at least according to this law is the case $x=1/2$,
so our theory is not dealing with corrections to the least likely event,
but rather corrections to the most likely events. This is because
of the fat tailed waiting times, which make the discussion of deviations
from familiar limiting laws a case study in its own right.
While the theory deals with most likely events, from the sampling point of view
these are still rare, as the probability of finding the occupation
time in a small interval close to the extremes of the arcsine law is still
small.

The organization of the paper is as
follows. In section II, we  outline the model and give
the necessary definitions. The behavior of the probability of
observing $N$ renewals in the interval $(0,t)$,  $p_N(t)$
is analyzed
in section III. In sections IV, V and VI, the densities of
the forward and backward
recurrence time, and the time interval straddling $t$,
denoted  $F$, $B$ and $Z$ respectively,
are derived.
In order to see the effects of the typical fluctuations and
large deviations, the fractional moments, e.g., $\langle F^q \rangle$,
 are considered and bi-fractal behavior is found.
In section VII, the behavior
of the occupation time  $T^+$ is  studied.
In the final section, we conclude the
paper with some discussions.
All along our work we demonstrate our results with numerical experiments
and compare between the statistical laws describing typical fluctuations
to those found here for the rare events.

\section{Model}\label{ldsect2}
Renewal process,  an idealized stochastic model for events that occur randomly in time, has a very rich and interesting mathematical structure and can be used as a foundation for building more realistic models \cite{Metzler2004The,Brokmann2003Statistic}. As mentioned, the basic mathematical assumption is that the time between the events are
IID random variables.
Moreover, renewal processes are often found embedded in other stochastic processes, most notably Markov chains.

\begin{figure}[htb]
  \centering
  \includegraphics[width=8cm, height=5.5cm]{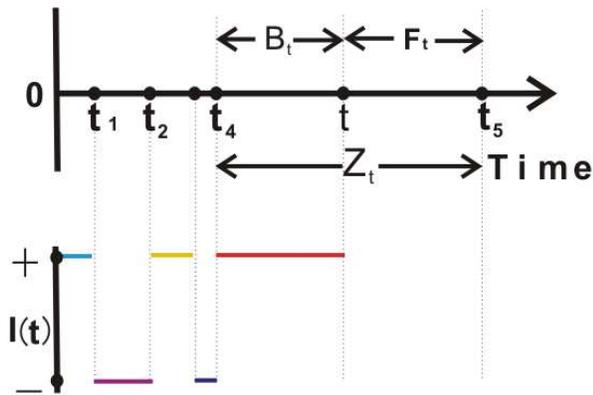}\\
  \caption{(color online) Illustration of a  renewal process.
The $t_i$ denotes  the time when the $i$-th event occurs.
$B_t$ and $F_t$ present the backward  and the forward recurrence time, respectively. In addition,  the time interval  straddling time $t$ is denoted with $Z_t$. The process $I(t)$, an alternating renewal process, is represented in the bottom of the figure. Here we suppose the initial state of the particle is  $+$.
We  see that the occupation time in the $+$ state is equal to $t_1+t_3-t_2+B_t$.}\label{rennew}
\end{figure}

Now, we  briefly outline the main ingredients of the renewal process \cite{Godreche2001Statistics}. It  is defined as follows: events occur at the random epochs of time $t_1$, $t_2$, $\ldots$, $t_N$, $\ldots$, from some time origin $t=0$.  When the time intervals between events, $\tau_1=t_1$, $\tau_2=t_2-t_1$, $\ldots$, $\tau_N=t_N-t_{N-1}$, $\ldots$, are IID random variables with a common PDF $\phi(\tau)$,  the process thus formed is a renewal process (see the top panel of the Fig.~\ref{rennew}).
We further consider the alternating renewal process  $I(t)$ in which the process alternates between $+$ and $-$ states.  A classical example is a Brownian motion $x(t)$ in dimension one, where we denote state $+$ with $x(t)>0$ and state $-$ for $x(t)<0$.
Generically, we  imagine that a device, over time, alternates between on and off states, like a  blinking dot \cite{Simone2005Fluorescence,Gennady2005Power}. Here we suppose the process starts in  $+$ state and stays in that state for a period of time $\tau_1$, then goes to $-$ state and remains for time $\tau_2$;  see bottom panel of the Fig.~\ref{rennew}. Clearly, it is natural to discuss the total time in state $+$ or $-$.
$T^+$ and $T^-$ are called  the occupation times in the $+$ and $-$ state, respectively and $T^++T^-=t$. For Brownian motion,  $\phi(\tau)\sim \tau^{-3/2}$ and the distribution of time in state $+$ is the well known arcsine law.

Motivated by  previous studies of  complex systems, we consider here PDFs with power law tails, i.e., for large $\tau$
\begin{equation*}\label{powerlawPDF}
  \phi(\tau)\sim \frac{\alpha \tau_0^\alpha}{\tau^{1+\alpha}}.
\end{equation*}
In this case, the first moment of $\phi(\tau)$ is divergent for $0<\alpha<1$. Here the index $\alpha>0$ and $\tau_0$ is a  time scale. As we show below, the {\reform full} form of $\phi(\tau)$ is of  importance  for the study of the large fluctuations.
An example is   the fat tailed PDF \cite{Metzler2000random}
\begin{equation}\label{ldeq3201111}
\phi(\tau)=\left\{
          \begin{split}
            &0, & \hbox{$\tau<\tau_0$;} \\
            &\alpha\frac{\tau_0^\alpha}{\tau^{1+\alpha}}, & \hbox{$\tau>\tau_0$.}
          \end{split}
        \right.
\end{equation}
Using the Tauberian theorem \cite{Feller1971introduction}, in Laplace space
\begin{equation}\label{ldeq3201}
  \widehat{\phi}(s)\sim 1-b_\alpha s^\alpha
\end{equation}
for small $s$,
where $s$ is conjugate to $\tau$, $b_\alpha=\tau_0^\alpha|\Gamma(1-\alpha)|$, and $0<\alpha<1$. In order to simplify the expression, we denote $\widehat{\phi}(s)$  as the Laplace transform of $\phi(\tau)$. When $1<\alpha<2$, the first moment  $\langle\tau\rangle=\int_0^\infty \tau\phi(\tau)d\tau$ is finite and the corresponding Laplace form \cite{Metzler2000random} is
\begin{equation}\label{ldeq3202}
  \widehat{\phi}(s)\sim 1-\langle\tau\rangle s+ b_\alpha s^\alpha
\end{equation}
for small $s$.
Notice that $\widehat{\phi}(0)=1$, which means that the PDF is normalized.
We would like to further introduce the one sided  L\'{e}vy distribution $\phi(\tau)=\ell_\alpha(\tau)$ with index $\alpha$, which is used in our simulations to generate  the process; see Appendix \ref{generatedpower}.
In Laplace space, one sided stable  L\'{e}vy distribution $\phi(\tau)$ is \cite{Metzler2004The}
\begin{equation}\label{ldeq32011}
 \int_0^\infty \exp(-s\tau)\phi(\tau)d\tau=\exp(-s^\alpha)
\end{equation}
and the small $s$ expansion is given by $\widehat{\phi}(s)\sim 1-s^\alpha$  with $0<\alpha<1$. For specific choices of $\alpha$, the closed form of the $\ell_\alpha(\tau)$ is tabulated for example in MATHEMATICA \cite{Burov2012Weak}.
In particular,  a useful special case is $\alpha=1/2$
\begin{equation}\label{ldeq32011levy}
  \ell_{1/2}(\tau)=\frac{1}{2\sqrt{\pi}}\tau^{-\frac{3}{2}}\exp\left(-\frac{1}{4\tau}\right).
\end{equation}
It implies that for large $\tau$,  $\ell_{1/2}(\tau)\sim \sqrt{4\pi}^{-1}\tau^{-3/2}$  so the first moment of the sojourn time diverges.
%
%

%
%
In the following we will draw on the research literature given by Godr\`{e}che and Luck \cite{Godreche2001Statistics}, which is  recommended for an introduction.
The number of renewal events  in the time interval between $0$ and time $t$ is
\begin{equation}\label{ldeq3101}
 N(t)=\max[N,t_N\leq t].
\end{equation}
Then we have the following relation $t_N=\tau_1+,\ldots, +\tau_N$. Now  we  introduce the forward recurrence time $F_t$, the time between $t$ and the next event
\begin{equation*}\label{ldeq3102}
  F_t=t_{N+1}-t;
\end{equation*}
see Fig.~\ref{rennew}.
While the corresponding backward recurrence time, the length between the last event before $t$ and the observation time $t$, is defined by
\begin{equation*}\label{ldeq31021}
  B_t=t-t_N.
\end{equation*}
Utilizing the  above two equations, we  get the time interval straddling time $t$, i.e., $Z_t$, which is
\begin{equation*}
  Z_t=B_t+F_t.
\end{equation*}
{\reform For simplification, we drop  the subscript, denoting the time dependence of the random quantities, from here on.}

\section{Number of renewals between $0$ and $t$} \label{ldpn}
We recap some of the basic results on the statistics of the number of renewal events.
The probability of the number of events $N$  up to time $t$ is
\begin{equation}\label{ldnu10111a}
  p_N(t)=\int_0^tQ_N(t')p_{0}(t-t')dt',
\end{equation}
and $Q_N(t')$ is the probability to have an event at time $t'$, defined by
\begin{equation}\label{ldnu10111aop}
  Q_N(t)=\int_0^t\phi(t')Q_{N-1}(t-t')dt'.
\end{equation}
Here $p_0(t)$ is the survival probability
\begin{equation}\label{spt}
  p_0(t)=\int_t^{\infty}\phi(y)dy,
\end{equation}
that is, the probability that the waiting time  exceeds the observation time $t$. For power law time statistics and large $t$,
\begin{equation*}
p_0(t)\sim \frac{b_\alpha}{|\Gamma(1-\alpha)|t^{\alpha}}.
\end{equation*}
Using  Eqs.~(\ref{ldnu10111a}, \ref{ldnu10111aop}) and convolution theorem leads to \cite{Montroll1965Random}
\begin{equation}\label{ldnu101}
  \widehat{p}_N(s)=\widehat{\phi}^N(s)\frac{1-\widehat{\phi}(s)}{s}
\end{equation}
with $N\geq 0$.
\subsection{Number of renewals between $0$ and $t$  with $0<\alpha<1$}
Rewriting Eq.~(\ref{ldnu101}), using the convolution theorem of Laplace transform,  and performing the inverse Laplace transform with respect to $s$, we  get a formal solution
\begin{equation}\label{ldnu102a}
  p_N(t)=\int_0^t\mathcal{L}^{-1}_\tau [\widehat{\phi}^N(s)]d\tau-\int_0^t\mathcal{L}^{-1}_\tau [\widehat{\phi}^{N+1}(s)]d\tau,
\end{equation}
where $N$ is a discrete
random variable and $\mathcal{L}^{-1}_{\tau}[\widehat{\phi}^N(s)]$  means the inverse Laplace transform, from the Laplace  space $s$ to real space $\tau$.



%
Summing the infinite series (summation over $N$),  the normalization condition $\sum_{N=0}^{\infty}p_N(t)=1$ is discovered as expected.
We  notice that Eq.~(\ref{ldnu102a}) can be further simplified when $\phi(\tau)$ is one sided L\'{e}vy distribution Eq.~\eqref{ldeq32011}. Then the inverse Laplace transform of Eq.~(\ref{ldnu102a}) gives
\begin{equation}\label{ldnu103}
p_N(t) =
\int_{t/(N+1)^{1/\alpha}}^{t/N^{1/\alpha}}\ell_\alpha(y)dy.
\end{equation}

As usual the large time limit is investigated with the small $s$ behavior of $\widehat{p}_N(s)$.
Utilizing Eq.~(\ref{ldeq32011}), the behavior of  Eq.~(\ref{ldnu101}) in the large $N$ limit and small $s$ is,
\begin{equation*}\label{ldnu105}
\begin{split}
 \widehat{\overline{ p}}_N(s) & \rightarrow b_\alpha s^{\alpha-1}\exp(-Nb_\alpha s^\alpha) \\
    &  =-\frac{1}{N\alpha}\frac{\partial}{\partial s}\exp(-Nb_\alpha s^\alpha).
\end{split}
\end{equation*}
Here note that $\int_0^\infty b_\alpha s^{\alpha-1}\exp(-Nb_\alpha s^\alpha)dN=1/s$. This means that with this approximation $N$ is treated as a continuous variables, which is fine since in fact we  consider a long time limit, and the limiting PDF of $N/t^\alpha$ is approaching a smooth function.
Hence,  we have  $\overline{ p}_N(t)$ to denote the continuous approximation.
%
First, using the property of  Laplace transform, i.e., $-\int_0^\infty \exp(-s\tau)\tau f(\tau)d\tau=\frac{\partial}{\partial s}\widehat{f}(s)$, secondly, performing the inverse Laplace transform on the above equation, we find  the well known result \cite{Aaronson1997introduction,Godreche2001Statistics,Schulz2014Aging}
\begin{equation}\label{ldnu106}
  \overline{p}_N(t)\sim\frac{t}{\alpha N^{1+1/\alpha}b_\alpha^{1/\alpha}}\ell_\alpha\Big(\frac{t}{(Nb_\alpha)^{1/\alpha}}\Big).
\end{equation}
Eq. (\ref{ldnu106}) is customarily called the inverse L\'{e}vy PDF.
Furthermore, using $\widehat{ \overline{p}}_N(s)$, we  find that $\widehat{\overline{p}}_u(t)$ can be expressed as Mittag-Leffler probability density
\begin{equation*}\label{ldnu106a}
\widehat{ \overline{p}}_{u}(t)=t^{\alpha-1}E_{\alpha,\alpha}(-ut^\alpha/b_\alpha),
\end{equation*}
where $\widehat{\overline{p}}_{u}(t)$ is the Laplace transform of $\overline{p}_{N}(t)$ with respect to $N$ and
a two-parameter function of the Mittag-Leffler type is defined by the series expansion \cite{Podlubny1999Fractional}
\begin{equation*}
  E_{\gamma,\nu}(z)=\sum_{n=0}^{\infty}\frac{z^n}{\Gamma(\gamma n+\nu)}
\end{equation*}
with $\gamma>0$ and $\nu>0$. {\reform Eq.~(14) describes statistics of functionals of certain  Markovian  processes, according to the Darling-Kac
theorem. It  was also investigated in the context of infinite ergodic theory \cite{Aaronson1997introduction} and continuous time random walks.}

\begin{figure}[htb]
 \centering
 \includegraphics[width=8cm, height=5.5cm]{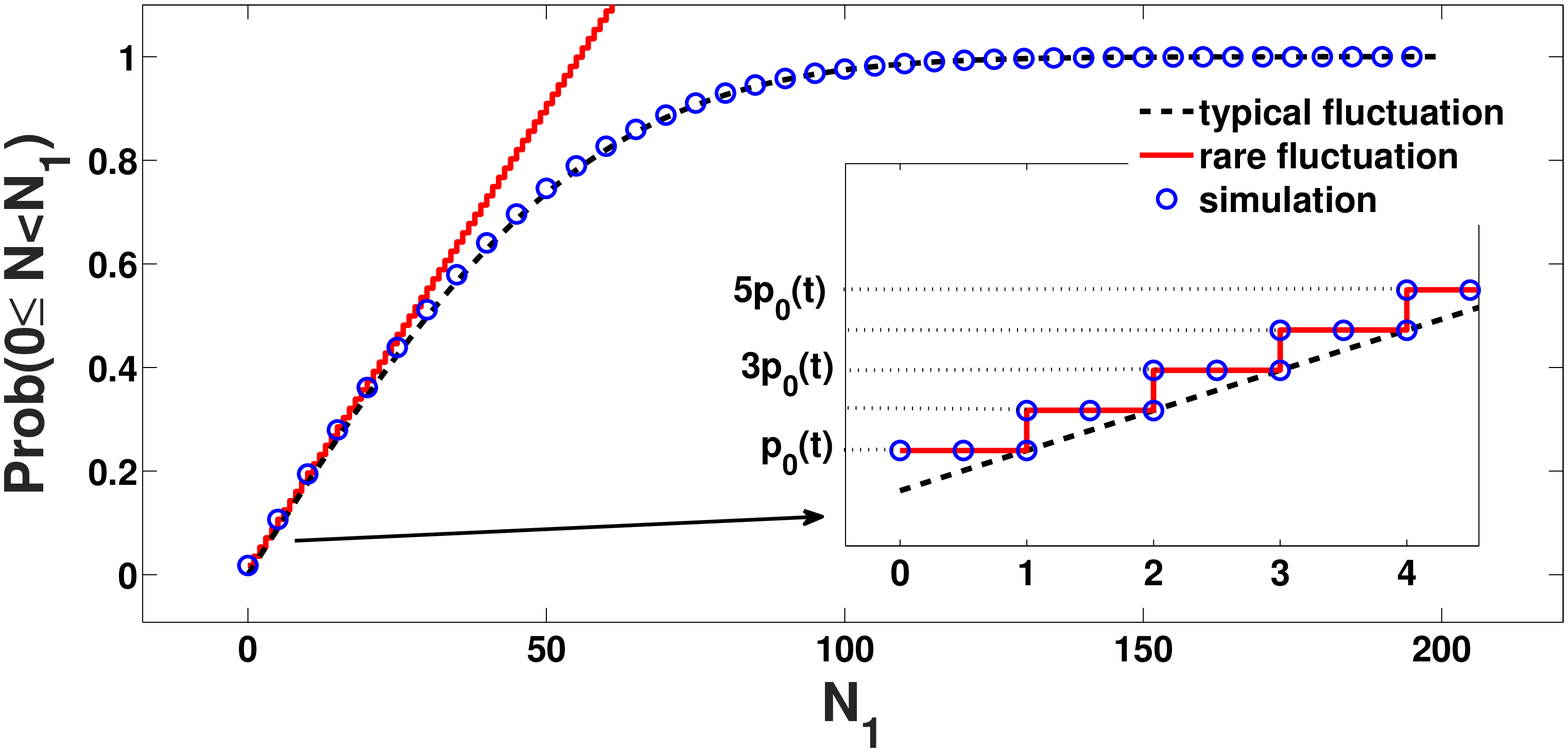}\\
 \caption{(color online) Comparison of analytical prediction Eq.~\eqref{ldnu104PRO1} (red solid line) for $Prob(0\leq N< N_1)$ with typical fluctuations  with $\alpha=1/2$.  We choose $t=1000$, waiting time PDF Eq.~(\ref{ldeq32011levy}), and $10^7$ trajectories. {\reform The typical fluctuations, plotted by dashed (black) lines, are obtained from Eq.~\eqref{ldnu106}.}
The rare  fluctuations are given by Eq.~\eqref{ldnu104PRO1} and they describe the probability very well for small $N_1$ (see inset).
}\label{eventJump}
\end{figure}

The well known limit theorem Eq.~(\ref{ldnu106}) is valid when $N$ and $t$ are large  and the ratio $N/t^\alpha$ is kept fixed. Now we consider rare events when $N$ is kept fixed and finite, say $N\sim 0$,~$1$,~$2$,~$3$ and $t$ is large. Using Eq.~(\ref{ldnu102a}) we find
\begin{equation}\label{ldnu104}
\lim_{t\rightarrow \infty}t^\alpha p_N(t) =\frac{b_\alpha}{\Gamma(1-\alpha)}.
\end{equation}
Note that $0<p_N(t)<1$ is  a probability, while $\overline{p}_N(t)$ is a PDF.  
To make a comparison between Eq.~\eqref{ldnu106} and Eq.~\eqref{ldnu104} we plot in Fig.~\ref{eventJump}, the probability that $N$ is in the interval $(0,N_1)$ versus $N_1$ and compare these theoretical  predictions to numerical simulations.  Integrating Eq.~\eqref{ldnu106} between $0$ and $N_1$, gives what we call the typical fluctuations.
While the result Eq.~\eqref{ldnu104} exhibits a staircase since according to this approximation
\begin{equation}\label{ldnu104PRO1}
\begin{split}
Prob(0\leq N<N_1)&\sim\sum_{N=0}^{floor[N_1]}\frac{b_\alpha}{\Gamma(1-\alpha)t^\alpha}\\
&\sim(floor[N_1]+1)p_0(t),
\end{split}
\end{equation}
where $floor[z]$
gives the greatest integer less than or equal to $z$.  From Fig.~\ref{eventJump} we see that,
besides the obvious discreteness of the probability, deviations between the two results can be considered marginal and non-interesting. Luckily this will change in all the examples considered below, as the statistical description of rare events deviates considerably from the known limit theorems of the field.

\subsection{Number of renewals between $0$ and $t$ with
 $1<\alpha<2$}
\begin{figure}[htb]
  \centering
  \includegraphics[
  width=9cm, height=6cm]{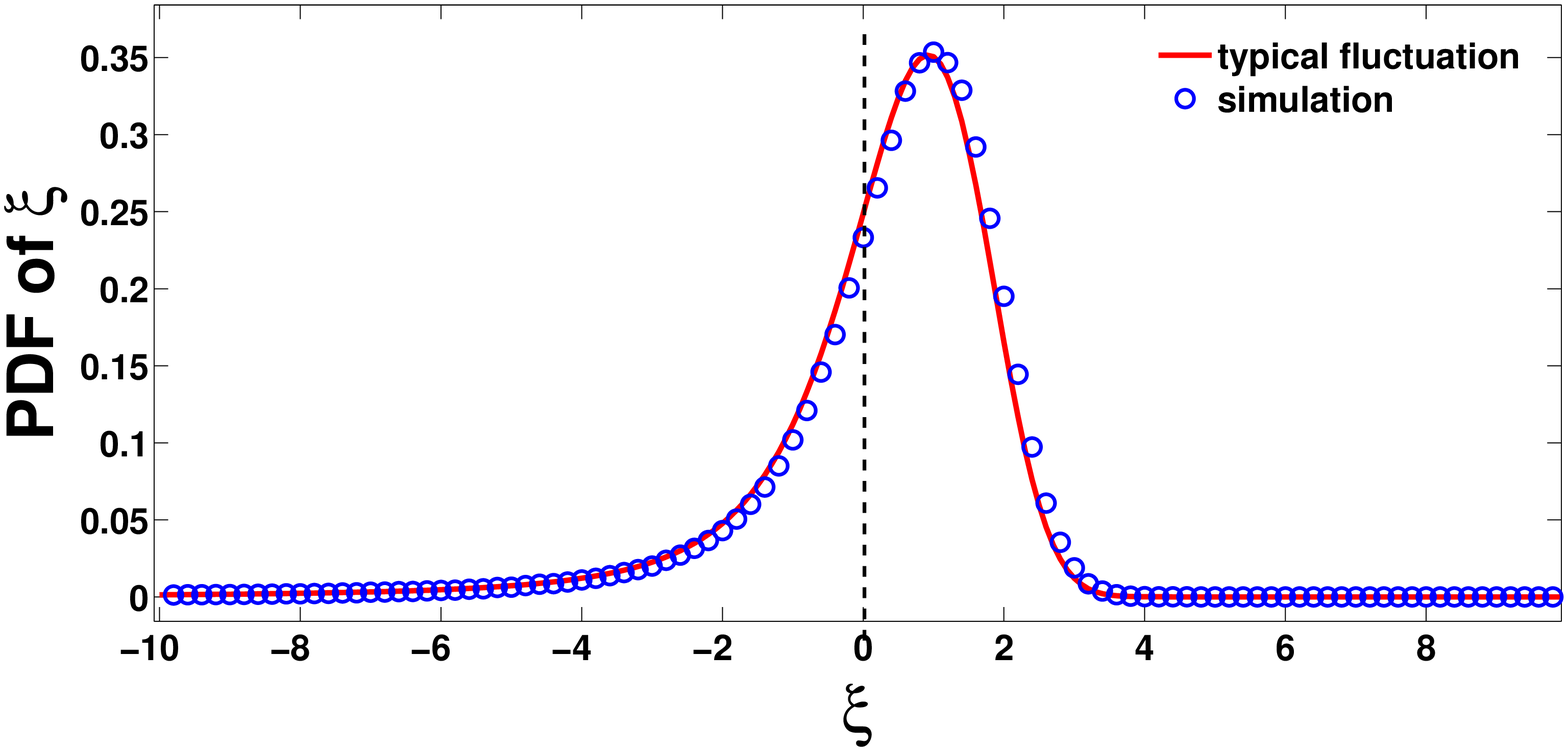}\\
  \caption{(color online) Simulation of the number of renewals using the rescaled variable $\xi=(N-t/\langle\tau\rangle)/(C_{ev}t^{1/\alpha})$. We show that the distribution of $\xi$ obtained from numerical simulations, converges to the L\'{e}vy density $L_{\alpha,1}(\xi)$ Eq.~(\ref{ldanu113}). This law describes the typical fluctuations, when $N-t/\langle \tau\rangle$ is of  the order of $t^{1/\alpha}$. For simulation we used $t=1000$, $3\times 10^6$ realizations, $\alpha=1.5$, and $\phi(\tau)$ given in Eq.~(\ref{ldeq3201111}).
%
%
%
As the figure shows, the right hand side of $L_{\alpha,1}(\xi)$ tends toward to zero rapidly.
 }\label{eventbulkalpha1.5}
\end{figure}

Based on Eq.~(\ref{ldnu101}), we  obtain a useful expression
\begin{equation}\label{ldanu111}
\widehat{\overline{p}}_u(s)=\frac{1-\widehat{\phi}(s)}{s}\int_0^\infty \exp(-uN+N\log(\widehat{\phi}(s)))dN.
%
%
%
\end{equation}
Here we consider the random variable, $\varepsilon=N-t/\langle\tau\rangle$, and explore its PDF denoted $\overline{p}_{\varepsilon}(t)$.  Applying Fourier-Laplace transform, $\varepsilon\rightarrow k$ and $t\rightarrow s$,  the PDF of $\varepsilon$ in Fourier-Laplace space is
\begin{equation}\label{ldanu111a}
\widehat{\overline{p}}_k(s)=\frac{1-\widehat{\phi}(s+\frac{ik}{\langle\tau\rangle})}{s+\frac{ik}{\langle\tau\rangle}}\frac{1}{-ik-\log(\widehat{\phi}(s+\frac{ik}{\langle\tau\rangle}))}.
\end{equation}
{\reform First, we consider  the limit of small $s$ and small $k$, {\bf and} the ratio $s/|k|^\alpha$ is fixed.
As we discuss below this leads to the description of what we call bulk or typical fluctuations, and these are described by standard central limit theorem. }
Substituting $\widehat{\phi}(s)$ into the above equation and taking inverse Laplace transform
\begin{equation}\label{ldanu112}
  \overline{p}_k(t)\sim \exp\Big(\frac{b_\alpha}{\langle\tau\rangle}\Big(\frac{ik}{\langle\tau\rangle}\Big)^\alpha t\Big).
\end{equation}
Fourier inversion of the above equation yields the PDF  $\overline{p}_{\varepsilon}(t)$,  written  in a scaling form \cite{Godreche2001Statistics}
\begin{equation}\label{ldanu113}
 \overline{p}_\varepsilon(t)\sim \frac{1}{C_{ev}t^{1/\alpha }}L_{\alpha,1}(\xi)
\end{equation}
with $C_{ev}=(b_\alpha/\langle\tau\rangle^{1+\alpha})^{1/\alpha}$ and $\xi=\varepsilon/(C_{ev}t^{1/\alpha})$; see Fig. \ref{eventbulkalpha1.5}. We  see that for fixed observation time $t$ the parameter $C_{ev}$ measures the PDF's width. Furthermore, the function $L_{\alpha,1}(x)$  is defined by
\begin{equation*}
\begin{split}
 L_{\alpha,1}(x) =\frac{1}{2\pi}\int_{-\infty}^{\infty}\exp(-ikx)\exp[(ik)^\alpha]dk,
\end{split}
\end{equation*}
where $L_{\alpha,1}(x)$ is the asymmetric L\'{e}vy PDF; see Appendix \ref{ldaphy11}.
Compared with the one sided L\'{e}vy distribution, $L_{\alpha,1}(x)$  holds two sides with the right hand side decaying rapidly. Moreover,  the second moment of $L_{\alpha,1}(x)$ diverges for $1<\alpha<2$.

{\reform As well known the central limit theorem (here of the L\'evy form) describes the central part of the distribution, but for finite though large $t$ it does not describe the rare events, i.e., the far tail of the distribution.  So far we investigated
the  typical or bulk statistics and as we showed they  are found for $N-t/\langle \tau\rangle \sim t^{1/\alpha}$.
Technically  this was obtained  using the  
 exact Laplace-Fourier transform, and then searching for a limit
where $s$ and $|k|^\alpha$ are small their ratio finite, as mentioned. However, it turns out that this limit is not unique. As we now show we can use the exact solution, assume both $k$ and $s$ are small, but their ratio $s/|k|$ finite and obtain a second meaningful solution. This in turn, leads to the description of rare events, i.e., the far tail of the distribution of the random variable $N$.
Roughly speaking, in this problem (and similarly all along the paper) we have two scales, one was just obtained and it grows like $t^{1/\alpha}$, the second  (with this example) is  $t/\langle \tau \rangle$, as we now show. This means that we have two ways to scale data, one emphasizing the bulk fluctuations (explained already) and the second the rare events.}

\begin{figure}[htb]
  \centering
  \includegraphics[
  width=9cm, height=6cm]{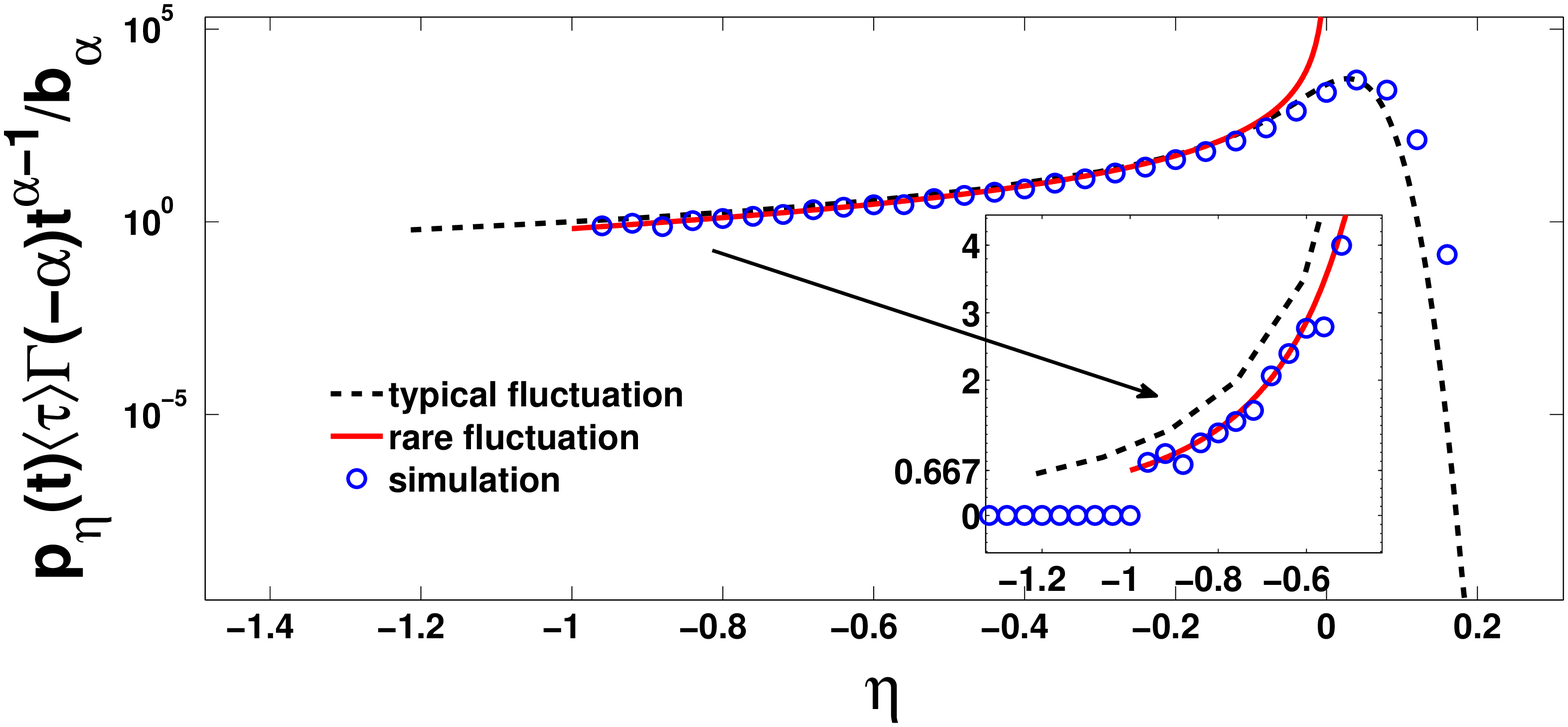}\\
  \caption{(color online) Simulation of renewal process with
  Eq.~\eqref{ldeq3201111}, yields $N$.
  We show how the scaled  PDF  of $\eta=(N-t/\langle\tau\rangle)/(t/\langle\tau\rangle)$, obtained from numerical simulations of the renewal process, converges to the non-normalized state Eq.~(\ref{ldanu1131}). The bulk description Eq.~\eqref{ldanu113} (dashed line) extends to $\eta<-1$, which is certainly not a possibility, and further it is not a valid approximation for $\eta\to -1$.
  Here we choose $t=1000$, $\alpha=1.5$ and $\tau_0=0.1$.  Deviations from typical fluctuations are clearly illustrated in the inset.}\label{eventforlargeN1.5}
\end{figure}

We now consider a second limiting law capturing the rare events valid when $\varepsilon$ is of the order of $t/\langle\tau\rangle$.
%
From Eq.~(\ref{ldanu111a}), we have
\begin{equation*}
  \widehat{\overline{p}}_k(s)\sim\frac{1}{s}-\frac{b_\alpha}{\langle\tau\rangle s}\Big(s+\frac{ik}{\langle\tau\rangle}\Big)^{\alpha-1}+\frac{b_\alpha}{\langle\tau\rangle s^2}\Big(s+\frac{ik}{\langle\tau\rangle}\Big)^{\alpha}.
\end{equation*}
Keep in mind that $s$ and $k$ are small and they are the same order. After performing inverse Fourier-Laplace transform, the asymptotic behavior of $\overline{p}_{\varepsilon}(t)$  is
\begin{equation}\label{ldanu1131gh}
  \overline{p}_{\varepsilon}(t)\sim -\frac{b_\alpha \alpha}{\Gamma(1-\alpha)}\Big(t(-\varepsilon\langle\tau\rangle)^{-\alpha-1}+\frac{1-\alpha}{\alpha}(-\varepsilon\langle\tau\rangle)^{-\alpha}\Big).
\end{equation}
This is the main result of this section. Here the above equation is only valid for negative $\varepsilon$.
We  see that  $\overline{p}_{\varepsilon}(t)$ decays like $t(-\varepsilon)^{-\alpha-1}$ for small negative $\varepsilon$.
Moreover, the scaling behavior of $\eta=\varepsilon\langle\tau\rangle/t$  yields
\begin{equation}\label{ldanu1131}
  \overline{p}_\eta(t)\sim \frac{b_\alpha  (-\eta)^{-\alpha-1}}{\langle\tau\rangle\Gamma(-\alpha)t^{\alpha-1}}\Big(1-\frac{1-\alpha}{\alpha}\eta\Big)
\end{equation}
with $-1<\eta<0$; see Fig.~\ref{eventforlargeN1.5}. It means that $\overline{p}_\eta(t)$ decays like $(-\eta)^{-\alpha-1}$ for  $\eta\rightarrow 0$, thus $\overline{p}_\eta(t)$ is not normalized.  Furthermore, for $\eta\rightarrow 0$, the dominating term $(-\eta)^{-1-\alpha}$ matches the left tail of Eq.~\eqref{ldanu113}; see Eq.~\eqref{ldappell103} in Appendix \ref{ldaphy11}.

For fixed observation time $t$, the central part of the PDF $\overline{p}_{\varepsilon}(t)$ is well illustrated by the typical fluctuations Eq.~(\ref{ldanu113}). While, its tail is described by  Eq.~(\ref{ldanu1131}), exhibiting the rare fluctuations. In order to discuss  the effect of typical fluctuations and large deviations,
we further consider the absolute moment of $\varepsilon$ \cite{Schulz2015Fluctuations}, defined by
\begin{equation}\label{ldanu11315}
  \langle|\varepsilon|^q\rangle=\int_{-\infty}^{\infty}|\varepsilon|^q \overline{p}_{\varepsilon}(t)d\varepsilon.
\end{equation}
Utilizing Eqs.~(\ref{ldanu113},~\ref{ldanu1131},~\ref{ldanu11315})
\begin{equation}\label{ldanu11311}
\langle|\varepsilon|^q\rangle\sim\left\{
  \begin{split}
  &(C_{ev})^qt^{q/\alpha}\int_{-\infty}^\infty |z|^qL_{\alpha,1}(z)dz, & \hbox{$q<\alpha$;} \\
  &\frac{b_\alpha q t^{q+1-\alpha}}{|\Gamma(1-\alpha)|\langle\tau\rangle^{q+1}(1+q-\alpha)(q-\alpha)}   , & \hbox{$q>\alpha$.}
  \end{split}
\right.
\end{equation}
Here we use the fact that $\int_{-\infty}^\infty |z|^qL_{\alpha,1}(z)dz$ is a finite constant for $q<\alpha$.
Note that to derive Eq.~(\ref{ldanu11311}) we use the non-normalized solution Eq.~(\ref{ldanu1131gh}) for $q>\alpha$, indicating that Eq.~(\ref{ldanu1131gh}) while not being a probability density, does  describe the high order  moments.
In the particular case $q=2$ (high order moment), we have $\langle|\varepsilon|^2\rangle=\langle(n-\frac{t}{\langle\tau\rangle})^2\rangle \sim\langle n^2\rangle-\langle n\rangle^2\sim\frac{2\tau_0^\alpha}{\langle\tau\rangle^3(\alpha-2)(\alpha-3)}t^{3-\alpha}$.
While this result is known \cite{Godreche2001Statistics}, our work shows that the second moment, in fact any moment of order $q>\alpha$, stems from the non-normalized density describing the rare fluctuations Eq.~\eqref{ldanu1131gh}. Other examples of such infinite densities will follow.

\begin{remark}
From simulations of the number of renewals $N$,
 Fig.~\ref{eventforlargeN1.5}, we see deviations from typical  results when $\eta>0$. As mentioned,
our theory covers the case $\eta<0$,
 so there is a need to extend the theory further.  Note that for $\eta>0$, the typical fluctuations  decay rapidly, while power law decay,
for intermediate values of $\eta$, is found on the left  (see Fig.~ \ref{eventforlargeN1.5}).
This intermediate power law behaviors can not continue forever, since $N\geq0$, and hence when $\eta\approx -1$ a new law emerges, Eq.~\eqref{ldanu1131}.
Possibly the large deviation principle can be used to investigate the case $\eta>0$.
\end{remark}



\section{The forward recurrence time}\label{LDsect21}
Several authors  investigated the distribution of $F$ both
for  $F\propto t$, meaning $F$ is of the  order of $t$, for $0<\alpha<1$ and also
$F \propto t^0$ for $\alpha>1$; see
Refs. \cite{Dynkin1961Selected,Feller1971introduction,Schulz2014Aging}.
These works
considered the typical  fluctuations of $F$, while we focus on the events of large deviations.
This means that we consider $F \propto t^0$ for $\alpha<1$ and $F \propto t$ for $1<\alpha$.
The forward recurrence time is an important topic of many stochastic processes, such as aging continuous time random walk processes (ACTRW) \cite{Schulz2014Aging,Kutner2017continuous}, sign renewals of Kardar-Parisi-Zhang Fluctuations \cite{Takeuchi2016Characteristic} and so on.     The forward recurrence time, also called the excess time (see  schematic Fig.~\ref{rennew}), is the time interval  between next renewal event and $t$.
In ACTRW, we are interested in the time interval that the particle has to wait before next jump if the observation is made at time t.
The PDF of the forward recurrence time is related to $Q_N(t)$ according to
\begin{equation}\label{ldf100}
  f_F(t,F)=\sum_{N=0}^{\infty}\int_0^t Q_{N}(\tau)\phi(t-\tau+F)d\tau;
\end{equation}
see  Eq.~(\ref{ldbasic102})
in Appendix \ref{ldbasic}.
In double Laplace space, the PDF  of $F$ \cite{Godreche2001Statistics} is
\begin{equation}\label{ldf101}
 \widehat{f}_{F}(s,u)=\frac{\widehat{\phi}(u)-\widehat{\phi}(s)}{s-u}\frac{1}{1-\widehat{\phi}(s)}.
\end{equation}
Based on the above equation, we will consider its analytic forms and  asymptotic ones.
In general case, the inversion of Eq.~(\ref{ldf101}) is a function that depends on  $F$ and $t$. While, for $\phi(\tau)=\exp(-\tau)$, the above equation can be simplified  as $f_F(t,F)=\exp(-F)$, which is independent of the observation time $t$. As expected, for this example we do not have an infinite density, neither multi-scaling of moments, since $\exp(-F)$ and more generally thin tailed PDFs,  do not have  large fluctuations like L{\'e}vy statistics.

%
%
\subsection{The forward recurrence time with $0<\alpha<1$}
\begin{figure}[htp]
    \begin{minipage}[t]{1\linewidth}
\begin{center}
 \includegraphics[height=6cm,width=9cm]{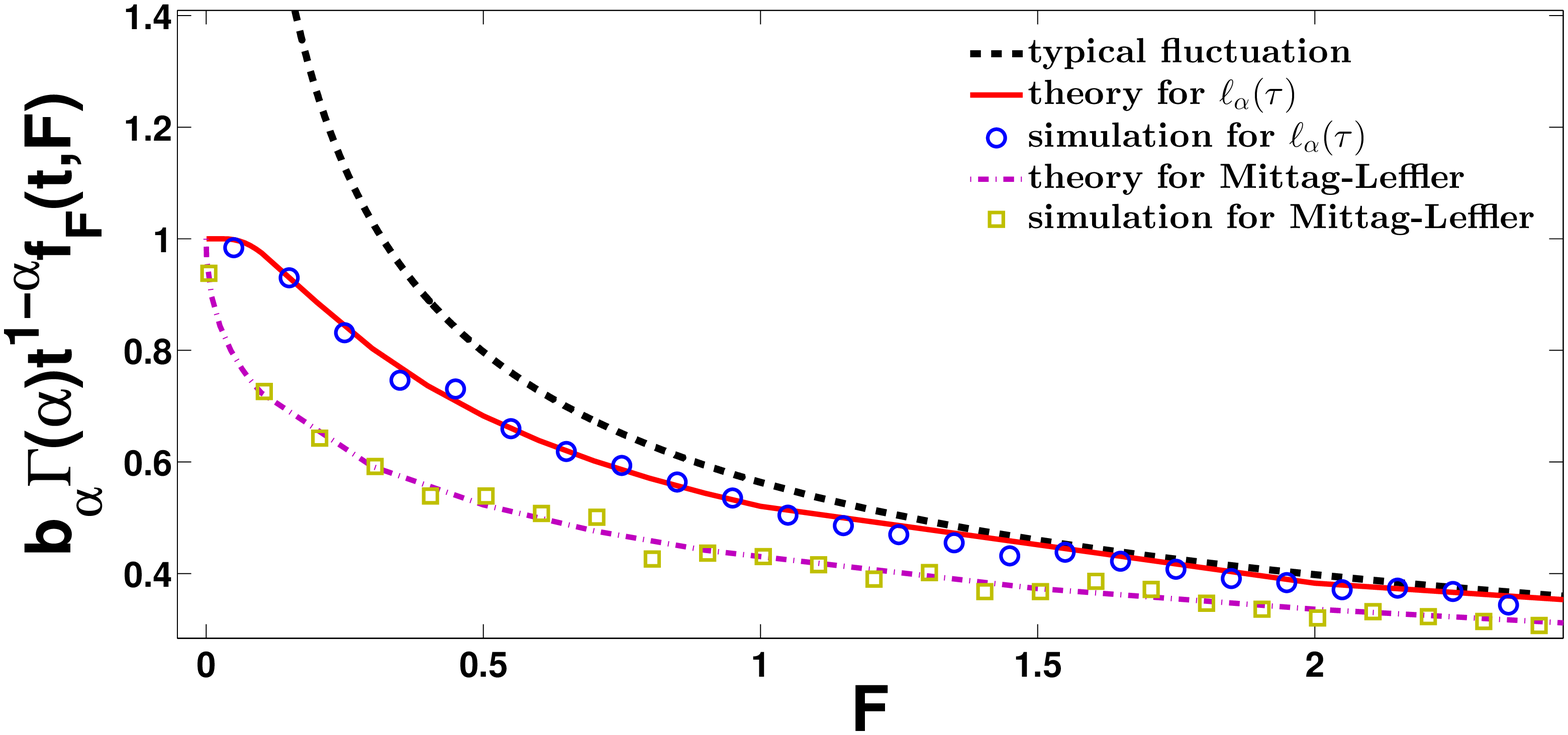}
\end{center}
    \caption{(color online) The behavior of  $f_{F}(t,F)$ for small $F$ with $\alpha=0.5$. The full (red) and the dash-dot (purple)  lines describing the large deviations are   the analytical results  Eqs.~(\ref{ldf103}) and \eqref{ldbeq102hii}, respectively.
The dashed (black) line given by Eq.~(\ref{ldf103c}) is Dynkin's limit theorem which gives the PDF  when $F$ is of the order of $t$, and $t$ is large.
Simulations are  obtained by averaging $10^7$ trajectories with $t=1000$.  Note that $b_\alpha\Gamma(\alpha)t^{1-\alpha}f_F(t,F)$ approaches to one for $F\rightarrow 0$. }
    \label{ForwardsmallF}
    \end{minipage}
    \begin{minipage}[t]{1\linewidth}
\begin{center}
    \includegraphics[height=6cm,width=9cm]{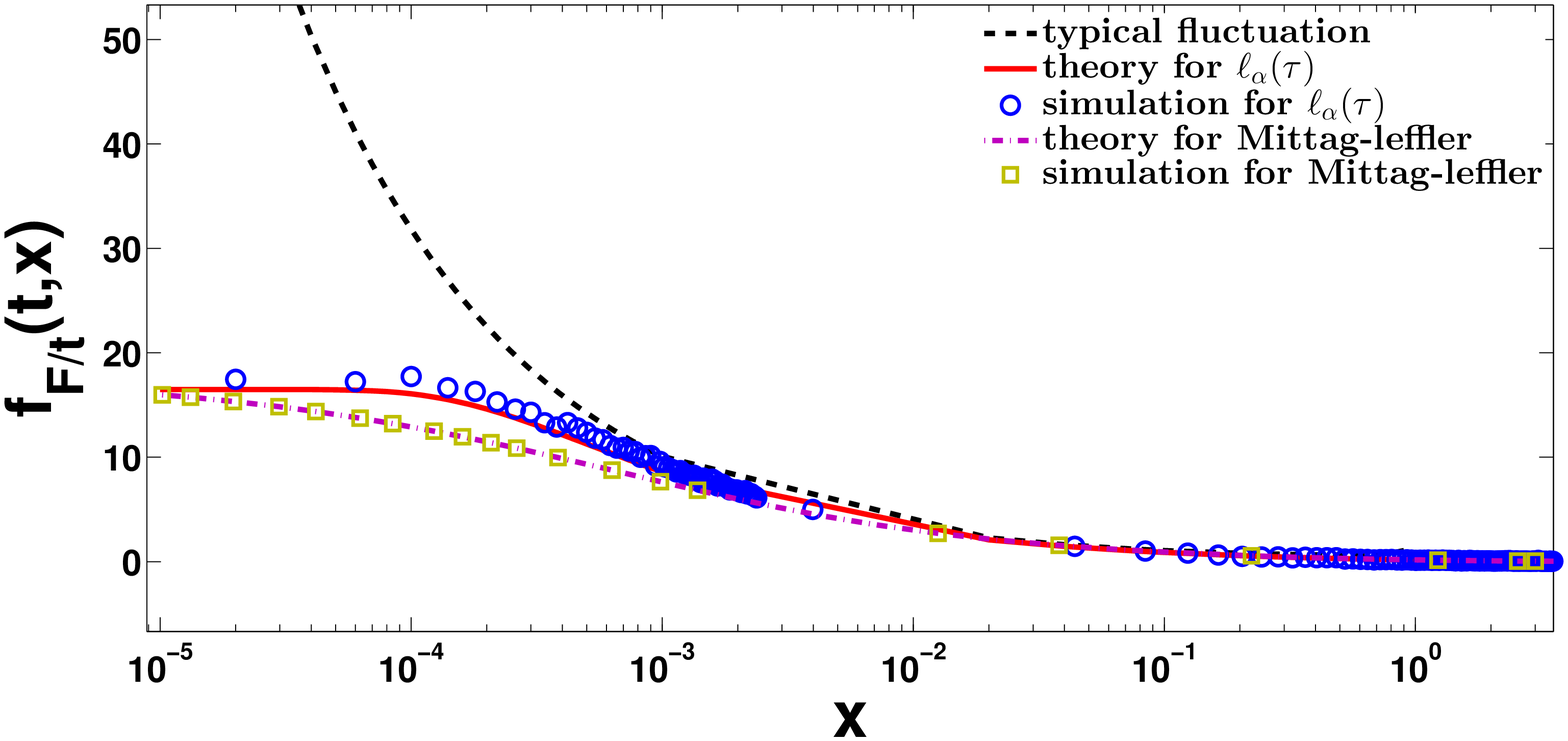}
    \caption{(color online) The behaviors of $f_{F}(t,F)$ with the scaling variable $x=F/t$ for $\alpha=0.5$ generated by $10^7$ trajectories with $t=1000$. The symbols  are the  simulation results.
  For $\alpha=0.5$, based on Eq.~(\ref{ldf103c}), we get $f_{F/t}(t,x)\sim(\pi(1+x)\sqrt{x})^{-1}$, which is shown by the dashed (black) line.  {\reform Here we use Eqs. \eqref{ldf103b}  and \eqref{ldf103bmlf} to predict theoretical results.}
   When $x\rightarrow 0$,  the PDF of $F$ depends on particular properties of $\phi(\tau)$, while for large $x$, the details of the PDF become non important, besides the value of $\alpha$. }
\label{Forwardscale}
\end{center}
\end{minipage}
\end{figure}

First, we are interested in the case of $F\ll t$. In Laplace space, this corresponds to $s\ll u$.
From Eq.~(\ref{ldf101})
\begin{equation}\label{ldf102}
  \widehat{f}_{F}(s,u)\sim \frac{1-\widehat{\phi}(u)}{u}\frac{1}{1-\widehat{\phi}(s)}.
\end{equation}
We notice that Eq.~(\ref{ldf102}) can be further simplified for a specific $\phi(\tau)$, namely Mittag-Leffler PDF \cite{Podlubny1999Fractional,Kozubowski2001Fractional}. In order to do so, we consider
\begin{equation}\label{ldbeq102h}
\phi(\tau)=\tau^{\alpha-1}E_{\alpha,\alpha}(-\tau^\alpha)
\end{equation}
with $0<\alpha<1$.
In Laplace space, $\widehat{\phi}(s)$ has the specific form
 \begin{equation}\label{ldbeq102hi}
\widehat{\phi}(s)=\frac{1}{1+s^\alpha}.
 \end{equation}
This distribution can be considered as the positive counterpart of Pakes's generalized Linnik distribution \cite{Jose2010Generalized} with the PDF having the form $(1+s^\alpha)^{-\beta}$, $0<\alpha<2$, $\beta>0$.
Plugging Eq.~(\ref{ldbeq102hi}) into Eq.~(\ref{ldf102}) leads to
\begin{equation*}
 \widehat{f}_F(s,u)\sim\frac{u^{\alpha-1}}{1+u^\alpha}\frac{1}{s^\alpha}.
\end{equation*}
Taking the  the double inverse Laplace transform yields
\begin{equation}\label{ldbeq102hii}
  f_F(t,F)\sim \frac{1}{\Gamma(\alpha)}E_{\alpha,1}(-F^\alpha)t^{\alpha-1};
\end{equation}
see Fig. \ref{ForwardsmallF}. Notice that $E_{\alpha,1}(0) =1$, so for $t>0$,
the PDF of $F$ for $F\rightarrow 0$ gives
$f_F(t,0)\sim t^{\alpha-1} /\Gamma(\alpha)$.

More generally, using Eq.~\eqref{ldeq3201}, we have
\begin{equation*}
 \widehat{f}_{F}(s,u)\sim\frac{1-\widehat{\phi}(u)}{u}\frac{1}{b_\alpha s^\alpha}.
\end{equation*}
Performing inverse double Laplace transform  leads to the main result of this section, and the density describing the large deviations is
\begin{equation}\label{ldf103}
  f_{F}(t,F)\sim \frac{\int_F^\infty\phi(y)dy}{\langle \tau^{*}\rangle},
\end{equation}
which exhibits interesting
aging effects \cite{Schulz2014Aging}.
Here $\langle \tau^*\rangle$ for large $t$ is equal to $(\Gamma(2-\alpha)\Gamma(1-\alpha))\int_0^t\tau^{'}\phi(\tau^{'})d\tau^{'}\sim b_\alpha\Gamma(\alpha)t^{1-\alpha}$, namely $\langle\tau^*\rangle$ is increasing with measurement time $t$,
and for  reasons that become clear later we may call it the effective average waiting time (recall the $\langle\tau\rangle$ is a constant only if $\alpha>1$).
The large deviations shows that  for large $F$  the forward recurrence time $f_F(t,F)$ decays as $F^{-\alpha}$. Furthermore,
 the integration of Eq.~\eqref{ldf103} over $F$ diverges since  $F^{-\alpha}$ is not integrable for large $F$.
 Hence Eq.~(\ref{ldf103}) is not a normalised density.
 For that reason, we may call $f_F(t,F)$ in Eq.~(\ref{ldf103}) an infinite density \cite{Rebenshtok2014Infinite}, the term infinite means non-normalizable, hence this is certainly not a probability density. Even though $f_F(t,F)$ Eq.~(\ref{ldf103}) is not normalized, it is used to obtain  certain observables, such as averages of observables integrable with respect to this non-normalized state. Besides, infinite densities play an important role in infinite ergodic theory \cite{Thaler2006Distributional,Akimoto2012Distributional} and intermittent maps \cite{Korabel2009Pasin}.

Using  Eq.~(\ref{ldf101}), we  find a formal solution to the problem
\begin{equation}\label{ldf103a}
f_F(t,F)=\phi(t+F)\ast_t\mathcal{L}^{-1}_t\Big[\frac{1}{1-\widehat{\phi}(s)}\Big].
\end{equation}
where  `$\ast_t$' is the Laplace convolution operator with respect to $t$ and the double  Laplace transform of  the function $f(t+F)$ is
\begin{equation*}
 \int_0^\infty \int_0^\infty \exp(-st-uF)f(t+F)dtdF=\frac{\widehat{f}(u)-\widehat{f}(s)}{s-u}.
\end{equation*}
We further discuss a special choice of $\phi(\tau)$, i.e., $\phi(\tau)=\ell_\alpha(\tau)$. After
some simple calculations, Eq.~(\ref{ldf103a}) gives
\begin{equation}\label{ldf103b}
\begin{split}
  f_F(t,F) & =\sum_{n=1}^{\infty}\frac{1}{n^{1/\alpha}}\int_0^t\ell_\alpha(t-\tau+F)\ell_\alpha\left(\frac{\tau}{n^{1/\alpha}}\right)d\tau \\
    & ~~~+\ell_\alpha(t+F).
\end{split}
\end{equation}
{\reform For Mittag-Leffler waiting time Eq.~\eqref{ldbeq102h}, we obtain
\begin{equation}\label{ldf103bmlf}
\begin{split}
 f_F(t,F)=&(t+F)^{\alpha-1}E_{\alpha,\alpha}(-(t+F)^{\alpha})\\
 &+\frac{1}{\Gamma(\alpha)}\int_0^t(\tau+F)^{\alpha-1}(t-\tau)^{\alpha-1}\\
 &\times E_{\alpha,\alpha}(-(\tau+F)^{\alpha})d\tau,
\end{split}
\end{equation}
}
from which we get the PDF of $x=F/t$ plotted in Fig.~\ref{Forwardscale}.

We now focus  on the typical fluctuations, namely
the case $F\propto t$ and both are large.
This means that $s$ and $u$ are small but of the same order.
Plugging  Eq.~(\ref{ldeq3201}) into  Eq.~(\ref{ldf101}), then taking double inverse Laplace transform,  leading to the normalized solution
\cite{Godreche2001Statistics,Dynkin1961Selected}
\begin{equation}\label{ldf103c}
  f_{F}(t,F)\sim\frac{\sin(\pi\alpha)}{\pi }\frac{1}{(\frac{F}{t})^\alpha(t+F)},
\end{equation}
which is plotted by the dashed (black) lines in Figs. \ref{ForwardsmallF} and \ref{Forwardscale}. The well known solution Eq.~(\ref{ldf103c}) describes the typical fluctuations when $F\sim t$.

To summarize, the forward recurrence time shows three distinct  behaviors: for $0<F\propto t^0$, the infinite density Eq.~\eqref{ldf103} rules, and only in this range, the PDF of  $F$ depends on the  behavior of $\phi(\tau)$; for $t^0\ll F\ll t$,  both Eqs.~\eqref{ldf103} and \eqref{ldf103c}  are valid and predict $f_F(t,F)\sim F^{-\alpha}$; for $F\gg t$, we use Eq.~\eqref{ldf103c} and then $f_F(t,F)\sim F^{-\alpha-1}$. Note that for certain observables, for example $B$ and $T^+$, when $B,T^+\to t$, their PDFs are also governed by the  shape  of $\phi(\tau)$; see below.

\subsection{The forward recurrence time with $1<\alpha<2$}
For $F\ll t$, according to Eq.~(\ref{ldf101})
\begin{equation*}\label{ldf104h01}
  \widehat{f}_F(s,u)\sim\frac{1-\widehat{\phi}(u)}{u\langle\tau\rangle s},
\end{equation*}
where as mentioned $\langle\tau\rangle$ is finite.
This can be finally inverted, yielding  {\reform the typical fluctuations} \cite{Feller1971introduction,Tunaley1974Asymptotic,Godreche2001Statistics}
\begin{equation}\label{ldf104h02}
  f_F(t,F)\sim \frac{1}{\langle\tau\rangle}\int_F^\infty \phi(y)dy.
\end{equation}
Since $1<\alpha<2$, Eq.~(\ref{ldf104h02}) is a normalized PDF and independent of the observation time $t$, which is different from Eq.~(\ref{ldf103}), but they have similar forms.  This is the reason why in the previous section we called $\langle\tau^*\rangle$ the
effective average waiting time.

\begin{figure}[htb]
  \centering
  \includegraphics[width=9cm, height=6cm]{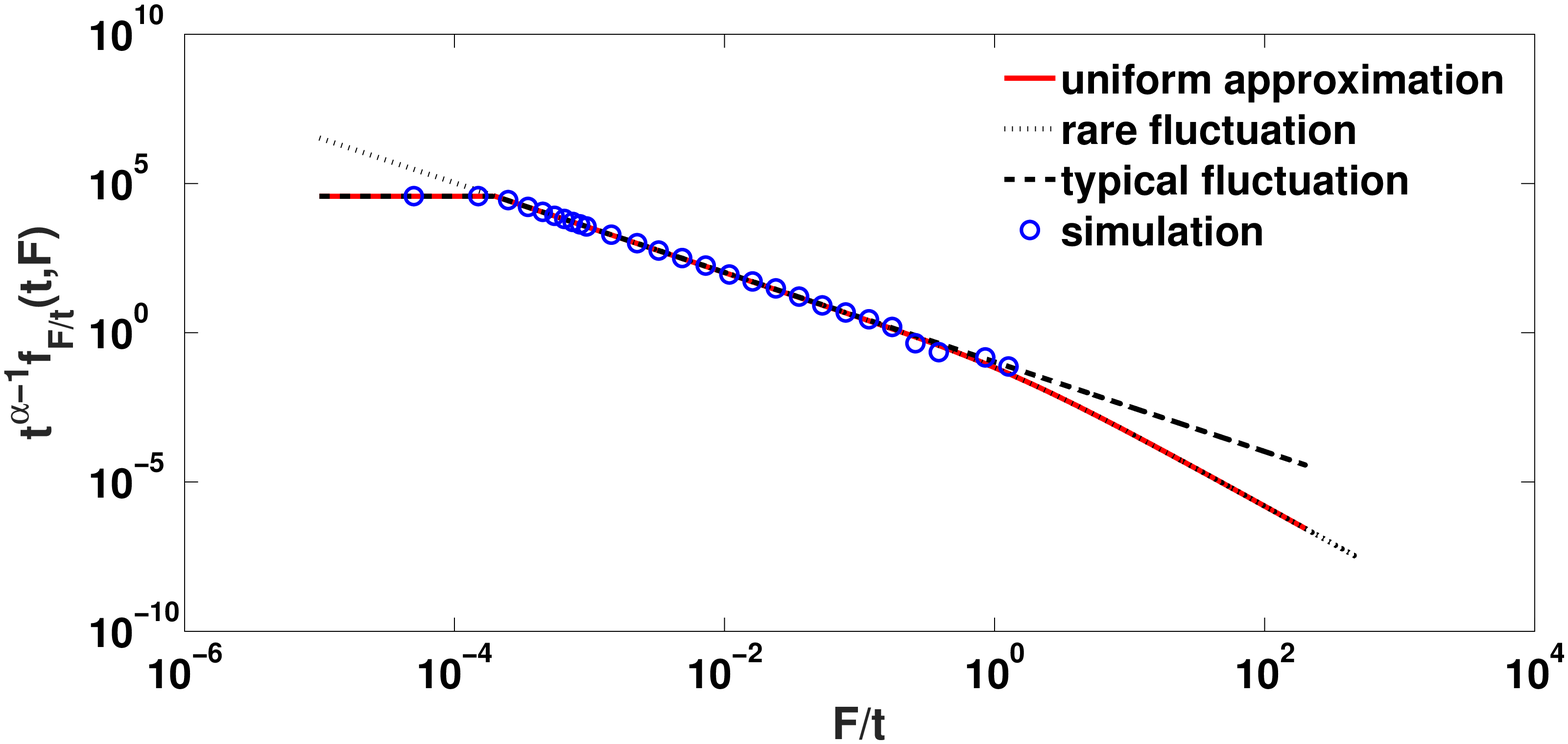}\\
  \caption{(color online) The PDF of $x=F/t$ multiplied by $t^{\alpha-1}$ versus $F/t$
  generated by $3\times10^6$ trajectories with Eq.~(\ref{ldeq3201111}). We choose $\alpha=1.5$, $\tau_0=0.1$ and $t=500$.  To obtain our theoretical results we used Eqs.~(\ref{ldf106}), (\ref{ldf107f}) and (\ref{ldf104h02}).
As the figure shows, Eq.~(\ref{ldf107f}), describing the large deviations is valid here for large $F/t$, though we experience a sampling problem in simulation (see remark).
%
%
 }\label{Forwardscale1.5}
\end{figure}

{\reform Next we discuss the uniform approximation, which is valid for varieties of $F$ and large $t$, namely within uniform approximation, we have the only condition that $t$ is large but the ratio of $F$ and $t$ arbitrary.}
It can be noticed that  Eq.~(\ref{ldf101}) can be arranged into the following formula
\begin{equation*}\label{ldf104a}
  \widehat{f}_{F}(s,u)=\frac{\widehat{\phi}(u)-1}{(s-u)(1-\widehat{\phi}(s))}+\frac{1}{s-u}.
\end{equation*}
For $F\neq t$, we may neglect the second term, then using $1-\widehat{\phi}(s)\sim \langle\tau\rangle s$ and inverting we get
\begin{equation}\label{ldf106}
f_{F}(t,F)   \simeq\frac{1}{\langle\tau\rangle} \int_F^{t+F}\phi(y)dy,
%
%
\end{equation}
which captures both the infinite density  and the bulk fluctuations;
see Fig.~\ref{Forwardscale1.5}.
Here, Eq.~(\ref{ldf106}) is true for large $t$ without considering the relation between $t$ and $F$. If $F\ll t$, Eq.~(\ref{ldf106}) can be approximated by Eq.~(\ref{ldf104h02}).

For the rare fluctuations, i.e., both $s$ and $u$ are small and comparable,  inserting Eq.~(\ref{ldeq3202}) into Eq.~(\ref{ldf101}), yields
\begin{equation*}\label{ldf107e}
\widehat{f}_F(s,u)\sim\frac{1}{s}+\frac{b_\alpha}{\langle\tau\rangle}\frac{u^\alpha-s^\alpha}{(s-u)s}.
\end{equation*}
For $F>0$, taking the double inverse Laplace transform, we find
\begin{equation}\label{ldf107f}
f_F(t,F)\sim\frac{b_\alpha}{|\Gamma(1-\alpha)|\langle\tau\rangle}(F^{-\alpha}-(F+t)^{-\alpha}),
\end{equation}
which is consistent with Eq.~(\ref{ldf104h02}) for large $F$ and $t\rightarrow \infty$. Besides, for $t\rightarrow \infty$, $f_F(t,F)$ decays as $F^{-\alpha}$ independent of the observation time $t$. On the other hand, if $F\gg t$, $f_F(t,F)$ grows linearly  with $t$, namely $f_F(t,F)\sim t/F^{-1-\alpha}$. In addition, using the asymptotic behavior of $\phi(F)$,  for large $F$  the uniform approximation Eq.~(\ref{ldf106}) reduces to Eq.~(\ref{ldf107f}). Still as for other examples in this manuscript, we may use Eq.~(\ref{ldf107f}) to calculate a class of high order  moments (for example $\alpha>q>\alpha-1$), i.e., those moments which are integrable with respect to this infinite density.

\begin{remark}\label{Ra1}
For simulations presented in Fig.~\ref{Forwardscale1.5}, we use $3\times10^6$ particles on a standard workstation, taking about $1$ day. We see that in this case we do not sample the rare events. In Ref. \cite{Rebenshtok2014Infinite}, simulations of the L\'{e}vy walk process with $10^{10}$ particles is performed, in order to explore graphically the far tails of  the  propagator of the  L\'{e}vy walk. When increasing the number of particles, we will observe rare events, {\reform however clearly in our case  $3\times 10^6$ 
realizations are simply  not sufficient for meaningful sampling. }
\end{remark}

With the help of the above equations, now we turn our attention to the fractional moments,   defined by
\begin{equation}\label{ldf108}
\langle F^q\rangle=\int_0^\infty F^qf_{F}(t,F)dF.
\end{equation}
Utilizing  Eq.~(\ref{ldf108}) and integration by parts, yields
\begin{equation}\label{ldf109}
  \langle F^q\rangle\sim\left\{
                       \begin{aligned}
                         &\frac{\int_0^{\infty}F^{q+1}\phi(F)dF}{(q+1)\langle\tau\rangle} , & \hbox{$q<\alpha-1$;} \\
                         &\frac{b_\alpha\Gamma(\alpha-q)\Gamma(1+q)t^{q+1-\alpha}}{\langle\tau\rangle|\Gamma(1-\alpha)|\Gamma(\alpha)(1+q-\alpha)}, & \hbox{$\alpha-1<q<\alpha$;} \\
                         &\infty, & \hbox{$q>\alpha$.}
                       \end{aligned}
                     \right.
\end{equation}
This is to say, for $q<\alpha-1$, $\langle F^q\rangle$ is a constant,  namely, it does not depend on the observation time $t$. Moments of order $q<\alpha$ are determined by the known result Eq.~(\ref{ldf104h02}), which describes typical fluctuations when $F$ is of the order of $t^0$. The rare fluctuations, described by Eq.~(\ref{ldf107f}), give information to events with $F\propto t$, and this non-normalized density Eq.~(\ref{ldf107f}) yields the moments of $\alpha-1<q<\alpha$; see  Eq.~(\ref{ldap1022})
in Appendix \ref{ldappfor1}.
Especially, if $q=1$, $\langle F\rangle\sim b_\alpha(\langle \tau\rangle\Gamma(3-\alpha))^{-1}t^{2-\alpha}$ so in this case the mean is determined by the infinite density.
When $q>\alpha$,  $\langle F^q\rangle$ is divergent. This is expected since the moment of order $q>\alpha$ of $\phi(\tau)$ diverges.

%
%
%
%
%
%
%
%
%

\section{The backward recurrence time}\label{LDsect22}
Compared with the forward recurrence time,  one of the important difference is that $B$  can not  be larger than $t$. In some cases, $B$ is  called the age at time $t$. Because in the lightbulb
lifetime example, it represents the age of the light bulb you find burning
at time t.
Similar to the derivation of  the forward recurrence time
\begin{equation*}
  f_B(t,B)=\sum_{N=0}^{\infty}\int_0^t Q_N(\tau)\delta(t-\tau-B)\int_B^{\infty}\phi(y)dyd\tau.
\end{equation*}
In Laplace space, let $t\rightarrow s$ and $B\rightarrow u$.
Using the convolution theorem of Laplace transform  and Eq.~(\ref{ldbasic102a}),
this gives
\begin{equation}\label{ldbeq101}
\widehat{f}_B(s,u)=\frac{1-\widehat{\phi}(s+u)}{s+u}\frac{1}{1-\widehat{\phi}(s)},
\end{equation}
which was  derived in Ref. \cite{Godreche2001Statistics}  using a different method.

\subsection{The backward recurrence time with $0<\alpha<1$}
\begin{figure}[htp]
    \begin{minipage}[t]{1\linewidth}
\begin{center}
 \includegraphics[height=6cm,width=9cm]{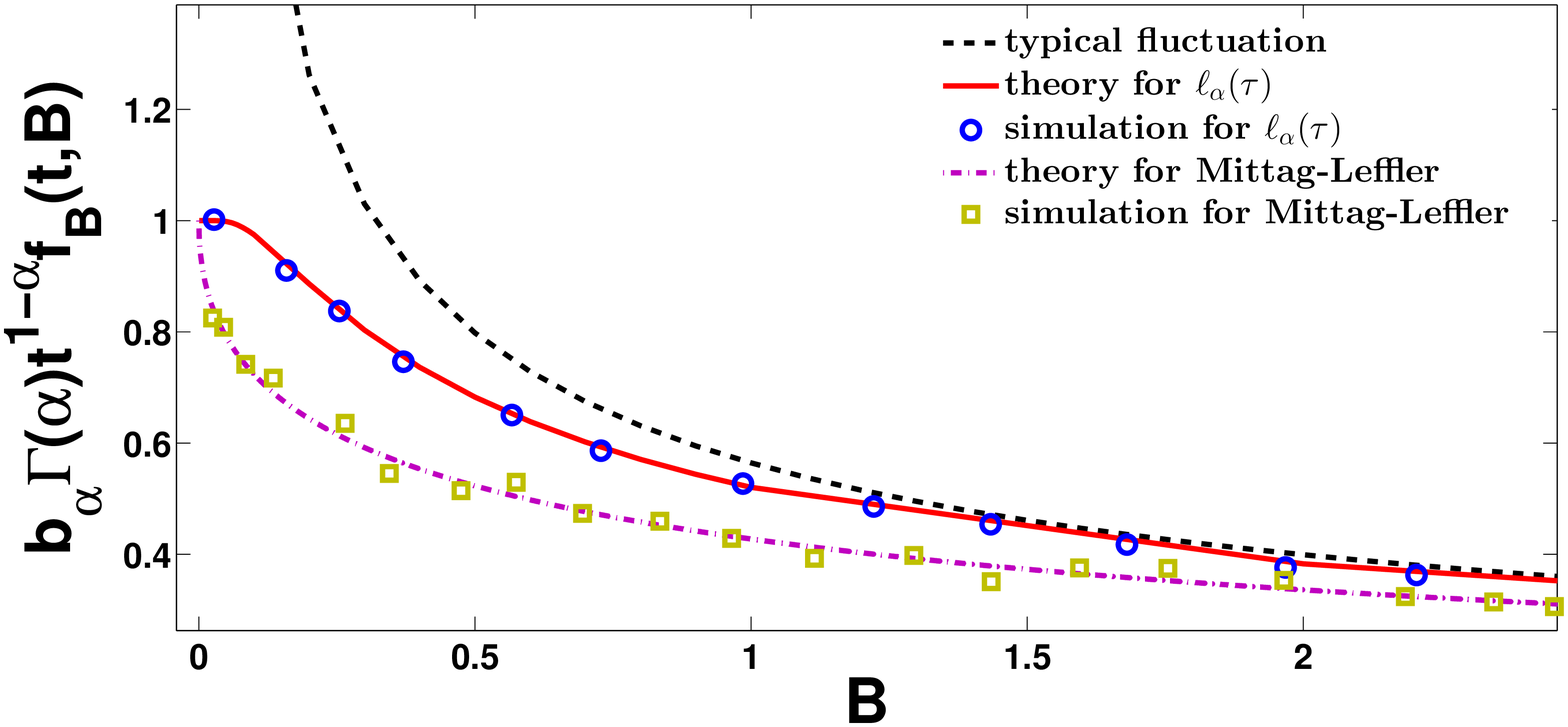}
\end{center}
    \caption{(color online) The scaled PDF of the backward recurrence time $B$, when $B$ is of the order of unity.
    The parameters are  $t=1000$ and $\alpha=0.5$. The full (red) and dash-dot lines are   the analytical result Eq.~(\ref{ldbeq103}) depicting the large deviations. For the typical result, we used Eq.~(\ref{ldbeq105a}), i.e., $f_B(t,B)\sim 1/(\pi\sqrt{B(t-B)})$. Besides, the symbols    are obtained by averaging $10^7$ trajectories with one sided L\'{e}vy stable distribution Eq.~(\ref{ldeq32011levy}) and Mittag-Leffler Eq.~(\ref{ldbeq102h}) time statistics, respectively.}
    \label{backwardsmallB}
    \end{minipage}
       \begin{minipage}[t]{1\linewidth}
\begin{center}
    \includegraphics[height=6cm,width=9cm]{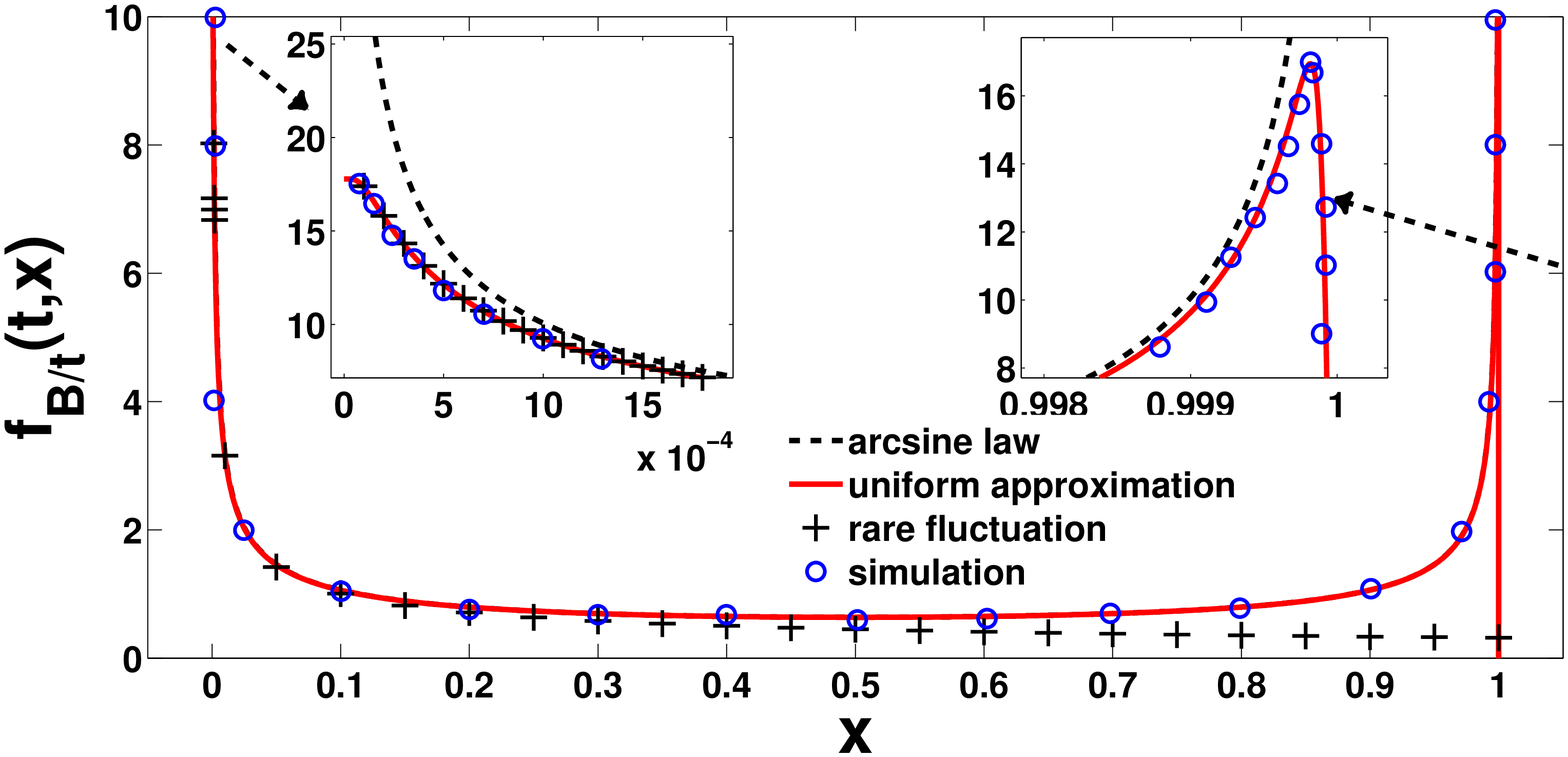}
    \caption{(color online)
 The PDF $f_{B/t}(t,x)$ versus  $x=B/t$ for a renewal process with $\phi(\tau)$ a one sided L\'{e}vy density
 Eq.~(\ref{ldeq32011levy}). Here we choose $t=1000$, and $\alpha=0.5$.  The dashed, the full lines and the symbols ($+$) present the arcsine law Eq.~\eqref{ldbeq105}, the analytical result Eq.~(\ref{ldbeqb103}), and rare events Eq.~(\ref{ldbeq103}), respectively.
 %
Notice that what the arcsine law predicts here is a symmetric distribution, while our results describing the large deviations exhibit non-symmetry. Furthermore our theory does not blow up at $x\rightarrow 0$ and $x\rightarrow1$, unlike the arcsine law.}
\label{backwardArcsin}
\end{center}
\end{minipage}
\end{figure}

First of all, we study the behaviors of large deviations.
For $B\ll t$, i.e., $s\ll u$
\begin{equation}\label{ldbeq102}
\widehat{f}_B(s,u)\sim \frac{1-\widehat{\phi}(u)}{u}\frac{1}{1-\widehat{\phi}(s)}.
\end{equation}
In the long time limit, i.e., $s\to 0$, $\widehat{f}_B(s,u)\sim (1-\widehat{\phi}(u))/(b_\alpha\,us^\alpha)$.
Performing the double inverse Laplace transform with respect to $s$ and $u$, respectively, yields
\begin{equation}\label{ldbeq103}
f_B(t,B)\sim \frac{\int_B^{\infty}\phi(y)dy}{\langle\tau^*\rangle},
\end{equation}
where $\langle\tau^*\rangle$ is defined below Eq.~\eqref{ldf103}.
It implies that $\lim_{B\rightarrow 0}f_B(t,B){\langle\tau^*\rangle}\sim 1$, which is confirmed in Fig.~\ref{backwardsmallB}.
Furthermore, note that $\int_0^\infty f_B(t,B)dB = \infty$,  which means that Eq.~(\ref{ldbeq103}) is non-normalized.
%
\begin{figure}[htb]
  \centering
  \includegraphics[width=9cm, height=6cm]{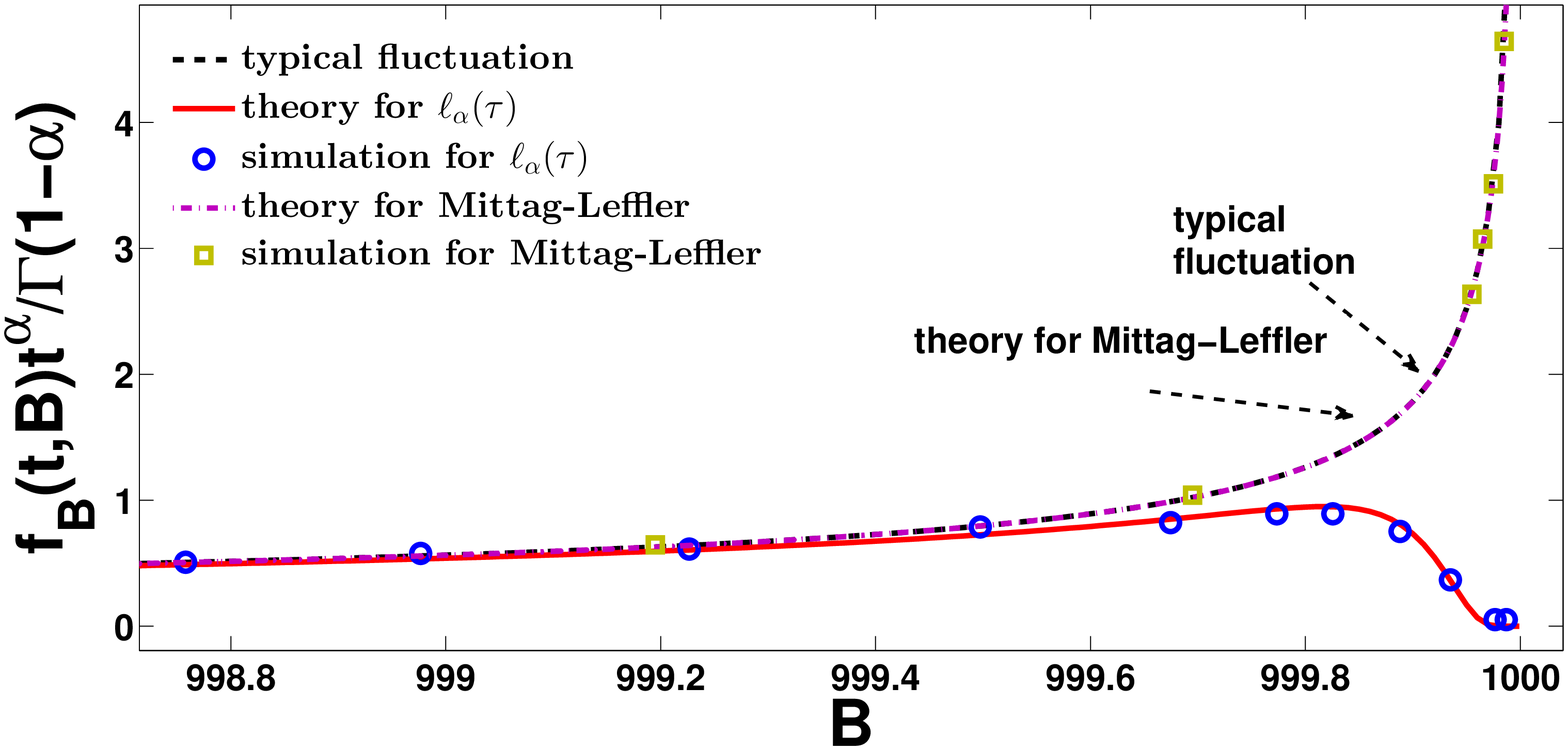}\\
  \caption{(color online) The relation between $f_{B}(t,B)$ and $B$ for large $B$. The parameters are  $t=1000$  and $\alpha=0.5$. For the typical result  we use Eq.~(\ref{ldbeq105}) and for the large deviations we use Eqs.~(\ref{ldbeqf10531o}) and (\ref{ldbeqf1056}).
The simulations, presented by symbols, are obtained by averaging $10^7$ trajectories. It is difficult to distinguish between the typical result and the theoretical result with Mittag-Leffler time statistics, while for the choice of $\phi(\tau)=\ell_{1/2}(\tau)$ Eq.~(\ref{ldeq32011levy}), the deviations are pronounced.
}\label{backwardlargeBAsymptotic}
\end{figure}

Now we  construct a uniform approximation, which is valid for a wider range of $B$, though $t$ is large. We rewrite Eq.~(\ref{ldbeq101}) as
\begin{equation*}\label{ldbeqlg101}
\widehat{f}_B(s,u)=\Big(\frac{1}{s+u}-\frac{\widehat{\phi}(s+u)}{s+u}\Big)\frac{1}{1-\widehat{\phi}(s)}.
\end{equation*}
For simplification, let $\phi(\tau)$ be the one sided L\'{e}vy stable distribution Eq.~\eqref{ldeq32011}.
Expanding the above equation, i.e., $1/(1-\widehat{\phi}(s))=\sum_{n=0}^{\infty}\widehat{\phi}^n(s)$, and   
then using the  convolution theorem of the Laplace transform
\begin{equation}\label{ldbeqb103}
\begin{split}
  f_B(t,B)  =&\delta(t-B)\int_t^{\infty}\ell_\alpha(y)dy+\Theta(t-B) \\
    &\times\int_B^{\infty}\ell_\alpha(y)dy\sum_{n=1}^{\infty}\frac{1}{n^{1/\alpha}}\ell_\alpha\left(\frac{t-B}{n^{1/\alpha}}\right);
\end{split}
\end{equation}
where $\Theta(z)$ represents the Heaviside theta function \cite{Polyanin2007Handbook}, which is equal to
 $0$ for $z<0$ and $1$ for $z>0$.
The  $\Theta(t-B)$ in Eq.~(\ref{ldbeqb103}) yields $B\leq t$ as expected.
In addition, for  $B\ll t$,  Eq.~(\ref{ldbeqb103}) reduces to Eq.~(\ref{ldbeq103}).
Note that, for $\alpha=1/2$, and comparable $t$ and $B$, Eq.~(\ref{ldbeqb103}) is consistent with the arcsine law, while, let $B$ go to either $0$ or $t$ (the extreme cases), the arcsine law does not work anymore; see Fig.~\ref{backwardArcsin}.

Now we turn our attention to the case of $B\to t$, using the random variable $\varepsilon=t-B\to 0$. In Laplace space,  the PDF of $\varepsilon$  is
\begin{equation}\label{ldbeqf1051}
\begin{split}
 \widehat{f}_{\varepsilon}(s,u_\varepsilon) & =\int_0^\infty\int_0^\infty\exp(-st-u_\varepsilon \varepsilon)f_{\varepsilon}(t,\varepsilon)dtd\varepsilon\\
    &=\widehat{f}_B(s+u_\varepsilon,-u_\varepsilon).
\end{split}
\end{equation}
According to Eq.~(\ref{ldbeq101})
\begin{equation}\label{ldbeqf1053}
\widehat{f}_\varepsilon(s,u_\varepsilon)=\frac{1-\widehat{\phi}(s)}{s}\frac{1}{1-\widehat{\phi}(s+u_\varepsilon)}.
\end{equation}
For $s\ll u_\varepsilon$, performing the  double inverse Laplace transform and using $B=t-\varepsilon$
\begin{equation}\label{ldbeqf10531}
f_B(t,B)=\int_t^\infty \phi(y)dy \mathcal{L}_{t-B}^{-1}\left[\frac{1}{1-\widehat{\phi}(u_\varepsilon)}\right].
\end{equation}
Let us consider a situation in which $\phi(\tau)$ is the Mittag-Leffler distribution
Eq.~(\ref{ldbeq102h}) with $0<\alpha<1$.
Next, plugging Eq.~(\ref{ldbeq102hi}) into Eq.~(\ref{ldbeqf10531}) yields
\begin{equation}\label{ldbeqf10531o}
  f_B(t,B)=p_0(t)\delta(t-B)+p_0(t)\frac{(t-B)^{\alpha-1}}{\Gamma(\alpha)}.
\end{equation}
It demonstrates that $t^\alpha f_B(t,B)$ decays like $(t-B)^{\alpha-1}$. Thus, if $B$ tends to the observation time $t$, we discover an interesting  phenomenon that $t^\alpha f_B(t,B) \to \infty$,  verified in Fig.~\ref{backwardlargeBAsymptotic}.
In general case, Eq.~(\ref{ldbeqf10531}) is not easy to calculate in real time exactly, though we use the  numerical inversion of  Laplace transform by MATLAB.  Expanding the above equation, we find
\begin{equation*}\label{ldbeqf1055}
f_\varepsilon(t,\varepsilon)= \int_t^{\infty}\phi(y)dy\left(\delta(\varepsilon)+\sum_{n=1}^{\infty} \mathcal{L}^{-1}_\varepsilon[\widehat{\phi}^n(u_\varepsilon)]\right).
\end{equation*}
Consider a specific $\phi(\tau)$, namely one sided L\'{e}vy stable distribution
\begin{equation}\label{ldbeqf1056}
f_B(t,B)= \int_t^{\infty}\ell_\alpha(y)dy\left(\delta(t-B)+\sum_{n=1}^{\infty}\frac{1}{n^{1/\alpha}}\ell_\alpha\Big(\frac{t-B}{n^{1/\alpha}}\Big)\right),
\end{equation}
which can be used for plotting.
To summarize, large deviations are observed for $B\propto t^0$ and $B\to t$, Eq.~(\ref{ldbeq103}) and Eqs.~(\ref{ldbeqf10531o}, \ref{ldbeqf1056}) respectively (see Fig.~\ref{backwardArcsin}), and these are non-symmetric for one sided L\'{e}vy distribution. Only when $B\sim t^0$, we find a non-normalized density, Eq.~(\ref{ldbeq103}).


Next we  discuss  the typical fluctuations when $B \propto t$. Combining  Eqs.~(\ref{ldeq3201}) and  (\ref{ldbeq101}), yields \cite{Godreche2001Statistics}
\begin{equation}\label{ldbeq105a}
 f_{B}(t,B)\sim\frac{\sin(\pi\alpha)}{\pi}\frac{1}{B^\alpha(t-B)^{1-\alpha}}\Theta(t-B).
\end{equation}
In a particular case $\alpha=1/2$, Eq.~(\ref{ldbeq105a}) reduces to the arcsine law $f_B(t,B)\sim(\pi\sqrt{B(t-B)})^{-1}$, which is plotted by the dashed (black) line in Figs. \ref{backwardsmallB}, \ref{backwardArcsin} and \ref{backwardlargeBAsymptotic}.
Let $x=B/t$, we  get a well known formula \cite{Godreche2001Statistics}
\begin{equation}\label{ldbeq105}
 f_{B/t}(x)\sim\frac{\sin(\pi\alpha)}{\pi}\frac{1}{x^\alpha(1-x)^{1-\alpha}}\Theta(1-x).
\end{equation}
In particular, for $\alpha=1/2$, Eq.~(\ref{ldbeq105}) reduces to the arcsine law \cite{Godreche2001Statistics} on $[0,1]$, see Fig. \ref{backwardArcsin}.

\subsection{The backward recurrence time with $1<\alpha<2$}
We again consider the limit $s\ll u$.
Combining Eqs.~(\ref{ldeq3202}) and (\ref{ldbeq101})
\begin{equation*}\label{ldbeq302}
\widehat{f}_B(s,u)\sim\frac{1-\widehat{\phi}(u)}{u}\frac{1}{\langle\tau\rangle s},
\end{equation*}
which, by the double inverse Laplace transform, yields the limiting result \cite{Feller1971introduction}
\begin{equation*}\label{ldbeq302a}
 f_B(t,B)\sim \frac{\int_B^{\infty}\phi(y)dy}{\langle\tau\rangle}.
\end{equation*}
If $B$ goes to $0$, $f_B(t,B)$ reduces to $1/\langle\tau\rangle$.

Now we turn our attention to the case when $B\to t$. According to Eq.~(\ref{ldbeqf1053})
\begin{equation}\label{ldbeq302aa}
f_B(t,B)\sim \int_{t}^{\infty}\phi(y)dy\mathcal{L}^{-1}_{t-B}\left[\frac{1}{1-\widehat{\phi}(u_\epsilon)}\right].
\end{equation}
For power law waiting time statistics, Eq.~(\ref{ldbeq302aa}) reduces to
\begin{equation*}
  f_B(t,B)\sim \frac{b_\alpha}{|\Gamma(1-\alpha)| t^\alpha}\mathcal{L}^{-1}_{t-B}\left[\frac{1}{1-\widehat{\phi}(u_\epsilon)}\right].
\end{equation*}
The inverse Laplace transform  gives  the limiting law when $B\to t$.

Let us proceed with the discussion of rare fluctuations. Substituting $\widehat{\phi}(u)$ and $\widehat{\phi}(s)$ into (\ref{ldbeq101}), leads to
\begin{equation*}\label{ldbeq302b}
\widehat{f}_B(s,u)\sim\frac{1}{s}-\frac{b_\alpha}{\langle\tau\rangle s}(s+u)^{\alpha-1}+\frac{b_\alpha}{\langle\tau\rangle }\frac{1}{s^{2-\alpha}},
\end{equation*}
when $s$ and $u$ are of the same order.
By inversion of   the above equation
\begin{equation}\label{ldbeq303}
\begin{split}
  f_B(t,B) & \sim \frac{b_\alpha}{\langle\tau\rangle |\Gamma(1-\alpha)|}B^{-\alpha}\Theta(t-B)
\end{split}
\end{equation}
with $B>0$. We  see that $f_B(t,B)$ blows up at $B\to 0$ and since $1<\alpha<2$ the solution Eq.~(\ref{ldbeq303}) is non integrable.

\begin{figure}[htb]
  \centering
  \includegraphics[width=9cm, height=6cm]{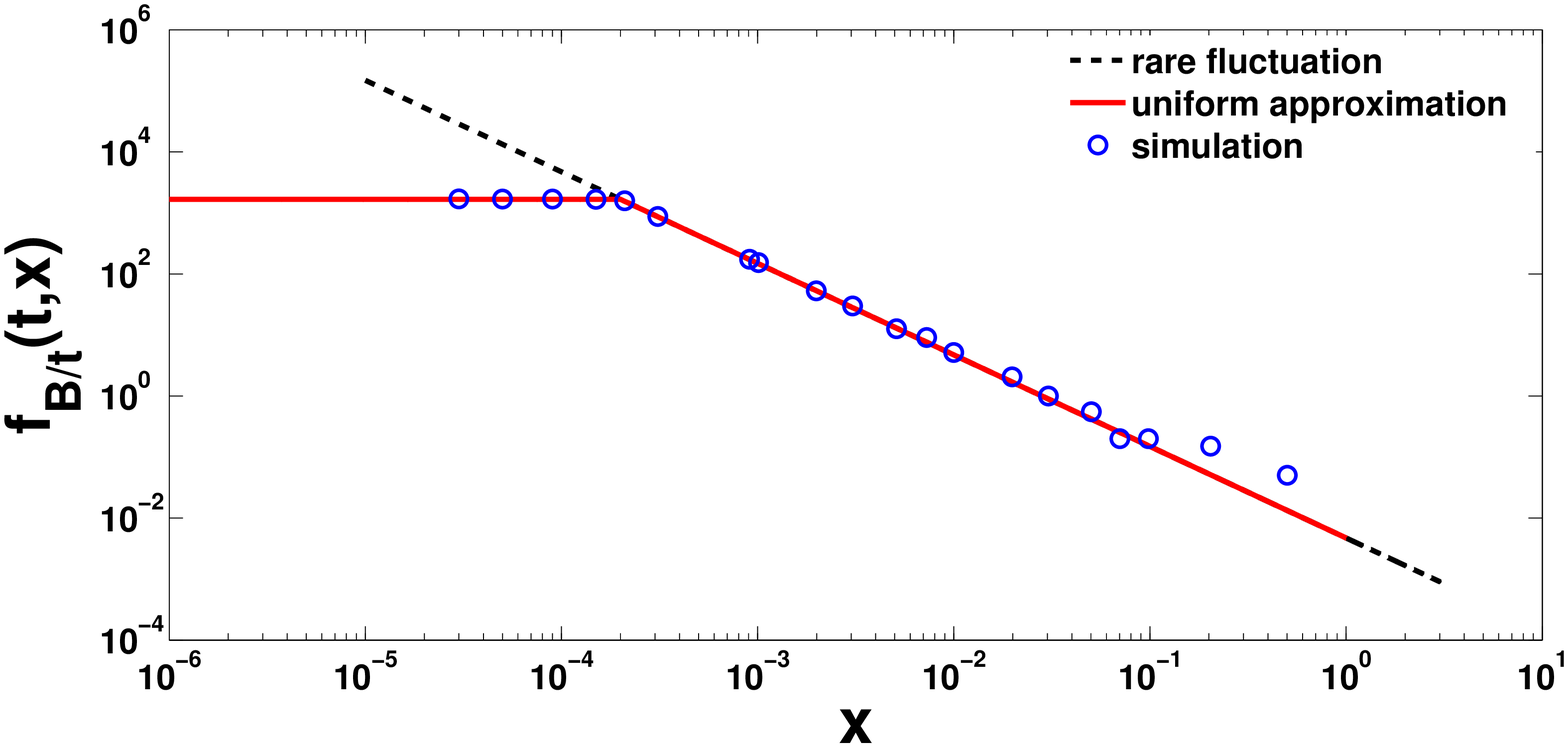}\\
  \caption{(color online)
  The PDF $f_{B/t}(t,B)$ versus the scaling variable $x=B/t$ for the waiting time PDF Eq.~(\ref{ldeq3201111}).
%
%
The parameters are  $t=500$, $\tau_0=0.1$  and $\alpha=1.5$. The solid (red) line is  the theory  Eq.~(\ref{ldbeq304}), and the dashed (black) line is Eq.~(\ref{ldbeq303}), which gives the PDF when $B$ is of the order of $t$, and $t$ is large. The simulations, presented by the symbols, are obtained by averaging $10^6$ trajectories.
}\label{backwardScale1.5}
\end{figure}

Next, utilizing Eq.~(\ref{ldbeq101}), the uniform approximation is
\begin{equation}\label{ldbeq304}
 f_B(t,B)\sim\frac{1}{\langle\tau\rangle}\Theta(t-B)\int_B^\infty \phi(y)dy.
\end{equation}
Note that $B$ is limited by the observation time $t$.
For large $B$, Eq.~(\ref{ldbeq304}) reduces to Eq.~(\ref{ldbeq303}).

The corresponding fractional moments are
\begin{equation}\label{ldbeq501}
  \langle B^q\rangle\sim\left\{
                       \begin{split}
                         &\frac{\int_0^\infty\phi(B)B^{q+1}dB}{\langle\tau\rangle(q+1)}, & \hbox{$q<\alpha-1$;} \\
                         &\frac{b_\alpha t^{q-\alpha+1}}{\langle\tau\rangle\left|\Gamma(1-\alpha)\right|(q-\alpha+1)},& \hbox{$q>\alpha-1$.}
                       \end{split}
                     \right.
\end{equation}
Since $B<t$ all moments are finite, unlike the case of the forward recurrence time.
The results show that the behaviors of fractional moments are divided into two parts. When $q<\alpha-1$, $\langle B^q\rangle$, is determined by the typical fluctuations and it is a constant. The rare fluctuations, described by Eq.~(\ref{ldbeq303}), give the information on events when $B\propto t$, and this non-normalized limiting law   gives the moments of $q>\alpha-1$.
The discussion of moments for $\alpha<1$ is given in Appendix \ref{qmentalpha0}.

\section{The  time interval  straddling  $t$}\label{sectstraddles}
%
%
%
The time interval straddling the observation time $t$ has been studied in Ref. \cite{Barkai2014From,Bertoin2006On,Chung1976Excursions,Getoor1979Excursions}, where some results about the typical fluctuations are announced and discussed. To consider general initial ensemble in  an
annealed transit time model \cite{Akimoto2016Distributional},  one  has to consider the time interval straddling time $t$ since the diffusion coefficient is governed by $Z$.
Based on the previous result \cite{Barkai2014From}, the PDF of $Z$ is given by the double Laplace inversion of
\begin{equation}\label{ldti101}
  \widehat{f}_Z(s,u)=\frac{1}{1-\widehat{\phi}(s)}\frac{\widehat{\phi}(u)-\widehat{\phi}(s+u)}{s},
\end{equation}
where $u$ is the Laplace pair of $Z$, and $s$ of $t$.
One important feature of
 $f_Z(t,Z)$ is the  discontinuity of its derivative at $Z=t$; see below.

\subsection{The  time interval  straddling $t$, $0<\alpha<1$}
\begin{figure}[htp]
    \begin{minipage}[t]{1\linewidth}
\begin{center}
 \includegraphics[height=6cm,width=9cm]{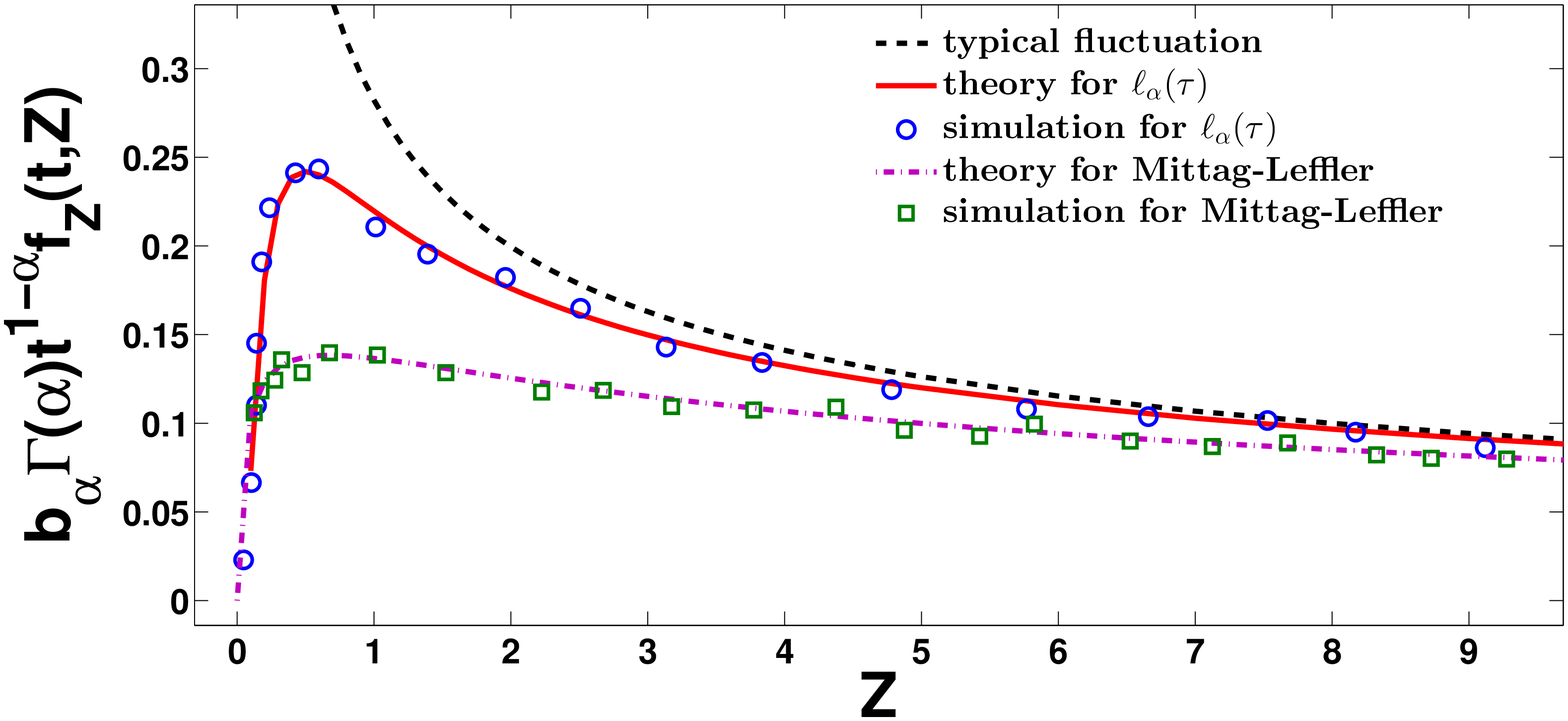}
\end{center}
    \caption{(color online)  Theory  and simulations of the statistical behaviors of the rare events of the time interval straddling time $t$,  with $\phi(\tau)$ Eqs.~(\ref{ldeq32011}) and (\ref{ldbeq102h}) for $t=1000$ and $\alpha=0.5$. The full (red)  and dash-dot (purple) lines are   theory  Eq.~(\ref{ldti103}), showing the large deviations and the corresponding simulation results are presented by symbols  obtained by averaging $10^{7}$ trajectories.}   
    \label{straddltimeSmallZ}
    \end{minipage}
    \begin{minipage}[t]{1\linewidth}
\begin{center}
    \includegraphics[height=6cm,width=9cm]{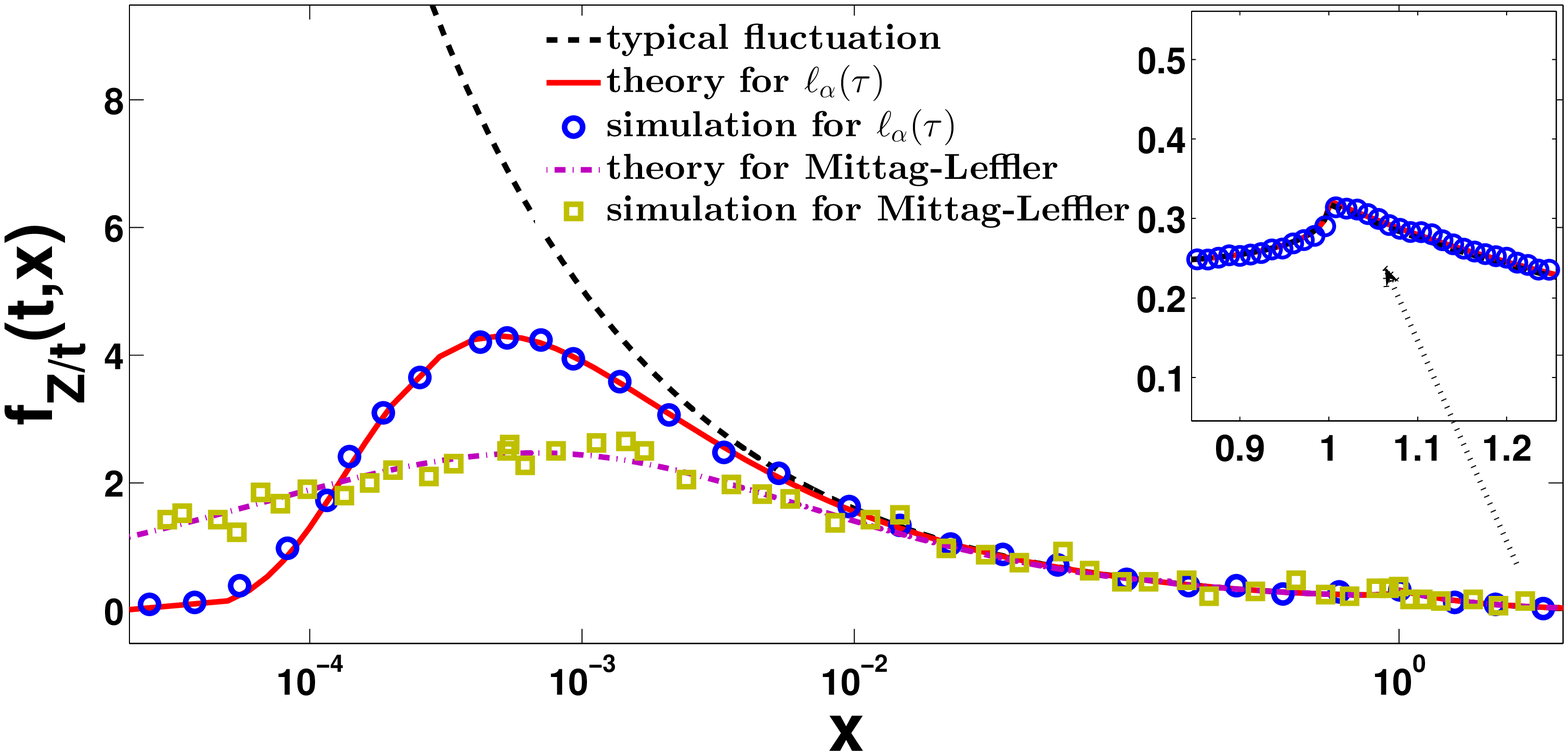}
    \caption{ (color online)
     The PDF of the rescaled variable $x=Z/t$ with $\alpha=0.5$.
The parameters are the same as in Fig.~\ref{straddltimeSmallZ}. {\reform The solid (red) and dash-dot (purple) lines present the theoretical results   Eqs.~(\ref{ldti103a}) and \eqref{mlfstraunir}, respectively.} Furthermore, the inset demonstrates that the first derivative of $f_Z(t,Z)$ is not continuous at $Z=t$.}\label{straddltimeTforZdt}
\end{center}
\end{minipage}
\end{figure}

Similar to previous sections, we first consider the  events of large deviations, namely, $Z$ is of the order of  $t^0$. Utilizing Eq.~(\ref{ldti101})
\begin{equation}\label{ldti102b}
 f_Z(t,Z)\sim\mathcal{L}^{-1}_t\left[\frac{1}{1-\widehat{\phi}(s)}\right]Z\phi(Z),
\end{equation}
which gives us an efficient way of calculation for $Z\ll t$.
In particular, combining Eqs. (\ref{ldeq3201}) and (\ref{ldti102b}), and taking  the inverse Laplace transform leads to
\begin{equation}\label{ldti103}
   f_Z(t,Z)\sim\frac{Z\phi(Z)}{\langle\tau^*\rangle},
\end{equation}
which is confirmed in Fig.~\ref{straddltimeSmallZ}.  Note that $\langle\tau^*\rangle$ is the same as that defined in Eq.~(\ref{ldf103}).
Keep in mind  that there is a  difference among small $F,B$ and $Z$. For small $Z$, $f_Z(t,Z)$ goes to $0$, while for Eqs.~(\ref{ldf103}) and (\ref{ldbeq103}) with $F, B\rightarrow 0$, $f_F(t,F)$ and $f_F(t,B)$ are equal to $1/\langle\tau^*\rangle$.
 In spite of these difference,  the asymptotic behavior of $f_Z(t,Z)$ is consistent with the PDF of the forward recurrence time and the  backward one with the increase of  $Z$.

We further consider the PDF of $Z$ more exactly. Taking the inverse Laplace transform of Eq.~(\ref{ldti101}) with respect to $u$ and $s$, respectively

\begin{equation}\label{ldti101c}
\begin{split}
  f_{Z}(t,Z) & =\phi(Z)\Big(\mathcal{L}^{-1}_{t}\Big[\frac{1}{s(1-\widehat{\phi}(s))}\Big] \\
    &~~~-\Theta(t-Z)\mathcal{L}^{-1}_{t-Z}\Big[\frac{1}{s(1-\widehat{\phi}(s))}\Big]\Big).
\end{split}
\end{equation}
In particular,
for a Mittag-Leffler density Eq.~(\ref{ldbeq102h}), the inversion of Eq.~(\ref{ldti101c}) can be further simplified as

\begin{equation}\label{mlfstraunir}
\begin{split}
  f_Z(t,Z) & =\frac{\phi(Z)}{\Gamma(1+\alpha)}\Big(t^\alpha-(t-Z)^\alpha \Theta(t-Z)\Big) \\
    &~~~+\phi(Z)(1-\Theta(t-Z)).
\end{split}
\end{equation}
It is interesting to note that Eq.~(\ref{mlfstraunir}) is a uniform approximation for Mittag-Leffler sojourn time. In addition,
we find that for $Z\ll t$  Eq.~(\ref{mlfstraunir}) reduces to Eq.~(\ref{ldti103}). On the other  hand,
when $Z>t$, the above equation yields $f_Z(t,Z)\sim t^\alpha\phi(Z)/\Gamma(1+\alpha)$. For $t\to 0$,  we see from Eq.~(\ref{mlfstraunir}) that  $\lim_{t\to 0}f_Z (t,Z)=\phi(Z)$  as expected.

Let us proceed with the discussion of a general waiting time PDF $\phi(\tau)$.
Expanding the term $(1-\widehat{\phi}(s))^{-1}$ of Eq.~(\ref{ldti101c})  in powers of $\widehat{\phi}(s)$, and then taking the inverse transform results in
 \begin{equation}\label{ldti103ag}
\begin{split}
  f_Z(t,Z) & =\phi(Z)\sum_{n=1}^{\infty}\Big(\int_0^t\mathcal{L}^{-1}_\tau[\widehat{\phi}^n(s)]d\tau-\Theta(t-Z)\\
   &~\times \int_0^{t-Z}\mathcal{L}^{-1}_\tau[\widehat{\phi}^n(s)]d\tau\Big)+\phi(Z)(1-\Theta(t-Z)).
\end{split}
\end{equation}
For the one sided L\'{e}vy stable distribution
\begin{equation}\label{ldti103a}
\begin{split}
  f_Z(t,Z) & =\ell_\alpha(Z)\sum_{n=1}^{\infty}\frac{1}{n^{1/\alpha}}\Big(\int_0^t\ell_\alpha\Big(\frac{\tau}{n^{1/\alpha}}\Big)d\tau\\
   &~~~-\Theta(t-Z)\int_0^{t-Z}\ell_\alpha\Big(\frac{\tau}{n^{1/\alpha}}\Big)d\tau\Big)\\
   &~~~~+\ell_\alpha(Z)(1-\Theta(t-Z)).
\end{split}
\end{equation}
Note that Eq.~(\ref{ldti103a}) is valid for all kinds of $t$ and $Z$. In Fig.~\ref{straddltimeTforZdt}, the scaling behaviors of $x=Z/t$ are displayed.
If $Z> t$,  Eq.~(\ref{ldti101c}) reduces to
\begin{equation}\label{ldti103ab}
f_{Z}(t,Z)\sim\frac{1}{\Gamma(1+\alpha)b_\alpha}t^\alpha\phi(Z).
\end{equation}

For $s$, $u$ small and comparable,  substituting $\widehat{\phi}(s)$ and $\widehat{\phi}(u)$ into Eq.~(\ref{ldti101}) yields
\begin{equation*}\label{ldti103b}
 \widehat{ f}_Z(s,u)\sim\frac{(s+u)^\alpha-u^\alpha}{s^{1+\alpha}},
\end{equation*}
and then taking the double inverse Laplace transform with respect to $u$ and $s$, gives the typical fluctuations \cite{Barkai2014From,Dynkin1961Selected}
\begin{equation}\label{ldti104}
f_Z(t,Z)\sim\frac{\sin(\pi\alpha)}{\pi} \frac{t^\alpha-(t-Z)^\alpha \Theta(t-Z)}{Z^{1+\alpha}},
\end{equation}
where $\Theta(x)$ =1 for $x\geq 0$  and is zero otherwise.
\subsection{The  time interval  straddling   time $t$ with $1<\alpha<2$}
\begin{figure}[htb]
  \centering
  \includegraphics[width=9cm, height=6cm]{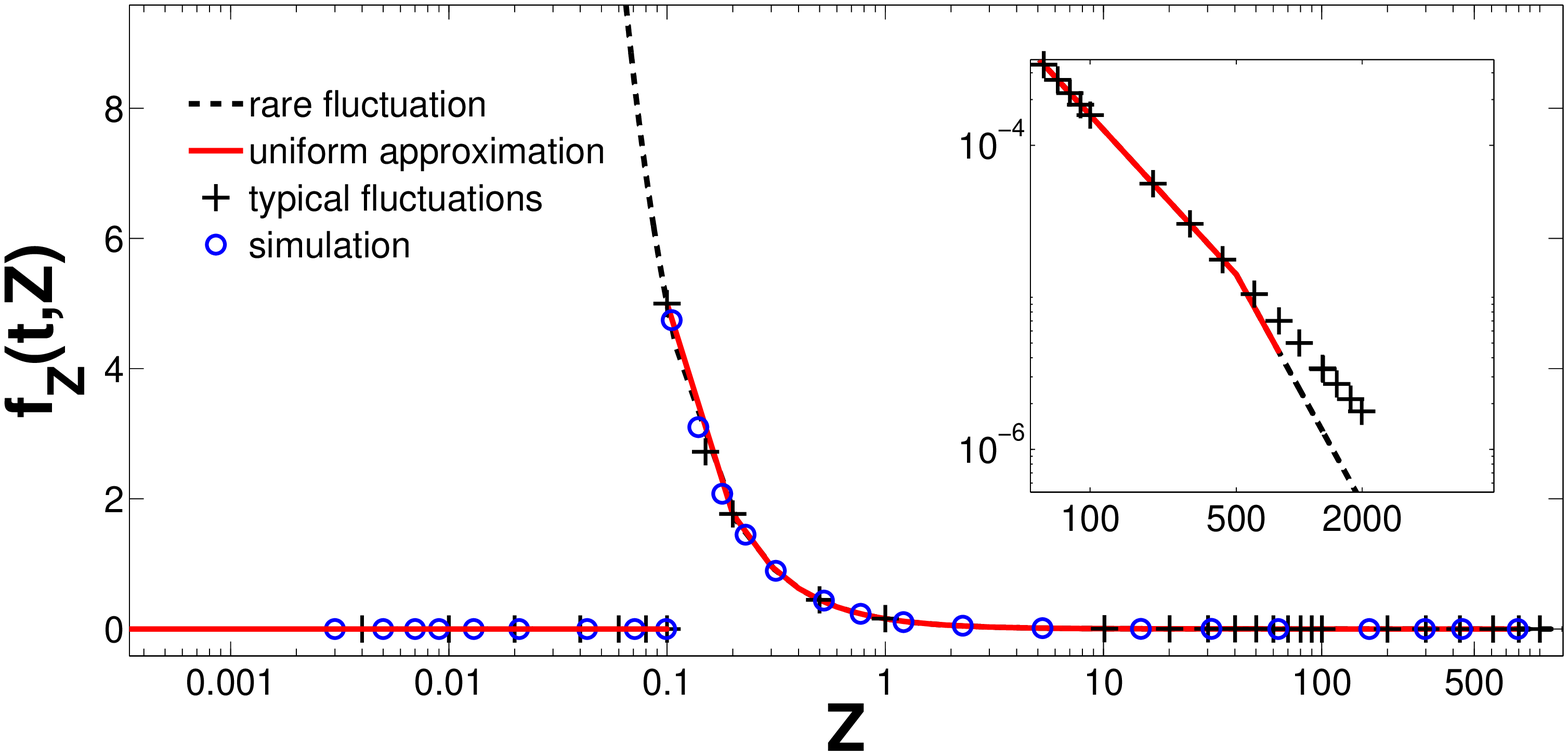}\\
 \caption{(color online) The PDF of the  straddling time $Z$ versus $Z$ for $\phi(\tau)$ Eq.~(\ref{ldeq3201111}). The parameters are the same as in Fig.~\ref{backwardScale1.5}. The rare fluctuations are given by Eq.~(\ref{ldsti105})
(the dashed (black) line), depicting the behaviors when $Z\propto t$. The solid (red) line is the uniform approximation  Eq.~(\ref{ldsti106b}). 
For the typical fluctuations, we use Eq.~\eqref{ldsti102} which is shown by the symbols  ($+$).
When $Z>t$, the rare fluctuations deviate from the typical fluctuations (see the inset).
}\label{straddltimeScale}
\end{figure}

For the typical fluctuations, i.e., $Z\sim t^0$. Based on Eq.~(\ref{ldti101}),
\begin{equation}\label{ldsti102}
f_Z(t,Z)\sim \frac{Z\phi(Z)}{\langle \tau\rangle};
\end{equation}
{\reform see Fig.~\ref{straddltimeScale}.}
Note that $f_Z(t,Z)$ tends to zero when $Z\to 0$.
%
We now discuss the rare fluctuations, i.e., $Z$ is  of the order of $t$. Plugging Eq.~(\ref{ldeq3202}) into Eq.~(\ref{ldti101}),
then performing the inverse Laplace transform, lead to
\begin{equation}\label{ldsti105}
 f_Z(t,Z)\sim\frac{b_\alpha Z^{-1-\alpha}}{\langle\tau\rangle\Gamma(-\alpha)}(t-(t-Z)\Theta(t-Z))
\end{equation}
with $Z>0$.
According to Eq.~(\ref{ldsti105}),  it gives us another representation of $f_Z(t,Z)$, namely
\begin{equation*}\label{ldsti106}
 f_Z(t,Z)\sim \left\{
                \begin{split}
                  &\frac{b_\alpha}{\langle\tau\rangle\Gamma(-\alpha)}Z^{-\alpha}, & \hbox{$Z< t$;} \\
                  &\frac{b_\alpha}{\langle\tau\rangle\Gamma(-\alpha)}tZ^{-\alpha-1}, & \hbox{$Z> t$.}
                \end{split}
              \right.
\end{equation*}

We now construct a uniform approximation that interpolates between
Eqs.~(\ref{ldsti102}) and (\ref{ldsti105}). We restart from Eq.~(\ref{ldti101}), but use Eq.~(\ref{ldeq3202}) only to approximate $1/(1-\widehat{\phi}(s))\sim 1/(\langle\tau\rangle s)$. After performing the double inverse Laplace transform, we arrive at
\begin{equation}\label{ldsti106b}
f_Z(t,Z)= \frac{C(t)}{\langle\tau\rangle}(t\phi(Z)-\Theta(t-Z)(t-Z)\phi(Z)),
\end{equation}
where we have added $C(t)=\langle\tau\rangle/(\int_0^t Z\phi(Z)dZ+\int_t^{\infty}t\phi(Z)dZ)$ as a  normalizing factor, satisfying $\lim_{t\rightarrow \infty}C(t)=1$.
In the long time limit, Eq.~(\ref{ldsti106b}) gives
\begin{equation}\label{ldsti103}
f_Z(t,Z)\sim\frac{t}{\langle\tau\rangle}\phi(Z)
\end{equation}
with $Z>t$.
It can be seen that Eq.~(\ref{ldsti103}) grows linearly with time $t$.

Similar to the calculations of $\langle F^q\rangle$ and $\langle B^q\rangle$
\begin{equation}\label{ldsti107}
 \langle Z^q\rangle\sim\left\{
                      \begin{split}
                       &\frac{1}{\langle\tau\rangle}\int_0^
                       \infty Z^{q+1}\phi(Z)dZ,&\hbox{$q<\alpha-1$;} \\
                        &\frac{b_\alpha t^{q-\alpha+1}}{\langle\tau\rangle\Gamma(-\alpha)(q-\alpha+1)(\alpha-q)}, &\hbox{$\alpha-1<q<\alpha$;} \\
                        &\infty, & \hbox{$\alpha<q$.}
                      \end{split}
                    \right.
\end{equation}
As expected  $\langle Z^0\rangle=1$. Similar to the previous examples,
when $\alpha-1<q<\alpha$, the moments
$\langle Z^q\rangle$ are obtained from Eq.~(\ref{ldsti105}), which is not a normalized PDF.
In particular, expanding the right hand side of Eq.~(\ref{ldti101})  to first order in $u$, and taking the inverse Laplace transform, lead to
$\langle Z\rangle\sim  (\langle\tau\rangle\Gamma(3-\alpha))^{-1}b_\alpha \alpha t^{2-\alpha}$,
which agrees with Eq.~(\ref{ldsti107}).

\section{Occupation time }\label{ldsec3}
The occupation time, the time spent by a process in a given subset of the state space during the interval of the observation, is  widely investigated  in mathematics and physics. It is a  useful quantity with a large number of  applications, for example the time spent by a one dimensional Brownian motion in half space, the time spent in the bright state for blinking quantum dot models \cite{Simone2005Fluorescence,Majumdar2002Local}, and the time that a spin occupies in a state up \cite{Majumdar1999Persistence}. 
Based on the alternating renewal process, here we focus on the study of the occupation time in the $+$ state.
In double Laplace space \cite{Godreche2001Statistics}
\begin{equation}\label{ldd102}
 \widehat{ f}_{T^+}(s,u)=\frac{2s+u}{2s(s+u)}+\frac{u(\widehat{\phi}(s+u)-\widehat{\phi}(s))}{2s(s+u)(1-\widehat{\phi}(s+u)\widehat{\phi}(s))};
\end{equation}
see the derivation of Eq.~(\ref{ldbasic105})
in Appendix \ref{ldbasic}.
In this model we start the process in the state up and down with equal probability.
Utilizing Eq.~(\ref{ldd102}) and taking $\epsilon= T^+-t/2$,
we  detect that the PDF  $f_{\epsilon}(t,\epsilon)$ is symmetric with respect to $\epsilon$ for a variety of $\phi(\tau)$.
As usual the difficulty is to find the solution in real time, namely find the PDF $f_{T^+}(t,T^+)$.
\begin{figure}[htp]
    \begin{minipage}[t]{1\linewidth}
\begin{center}
 \includegraphics[height=6cm,width=9cm]{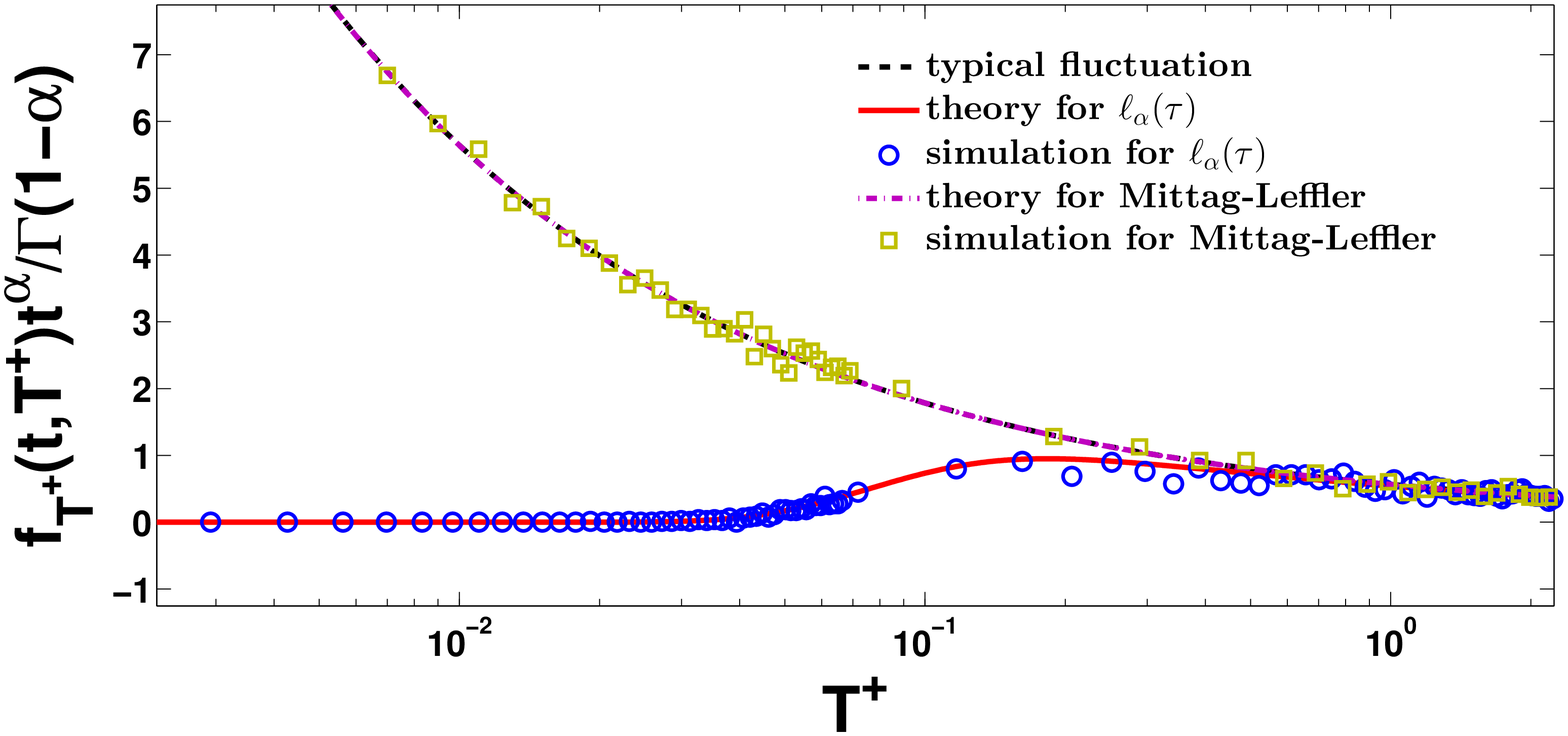}
\end{center}
\setlength{\abovecaptionskip}{0pt}
\setlength{\belowcaptionskip}{0pt}
    \caption{(color online) The scaled PDF of the occupation time  versus $T^{+}$ generated by the trajectories of particles with $\alpha=0.5$ and $t=1000$ for $T^+\ll t$.
The solid (red) line and the dash-dot line correspond to the theoretical results given by Eqs.~(\ref{ldd107}) and (\ref{ldd10603}), respectively, depicting the large deviations with $T^+\propto t^0$. The dashed (black) line, given by Eq.~(\ref{ldd104}),  shows the typical fluctuations.  Note that it overlaps with theoretical result of Mittag-Leffler waiting time, the top curve in the figure.
}   
    \label{occpationSmallTplus}
    \end{minipage}
   \begin{minipage}[t]{1\linewidth}
\begin{center}
\setlength{\abovecaptionskip}{0pt}
\setlength{\belowcaptionskip}{0pt}
    \includegraphics[height=6cm,width=9cm]{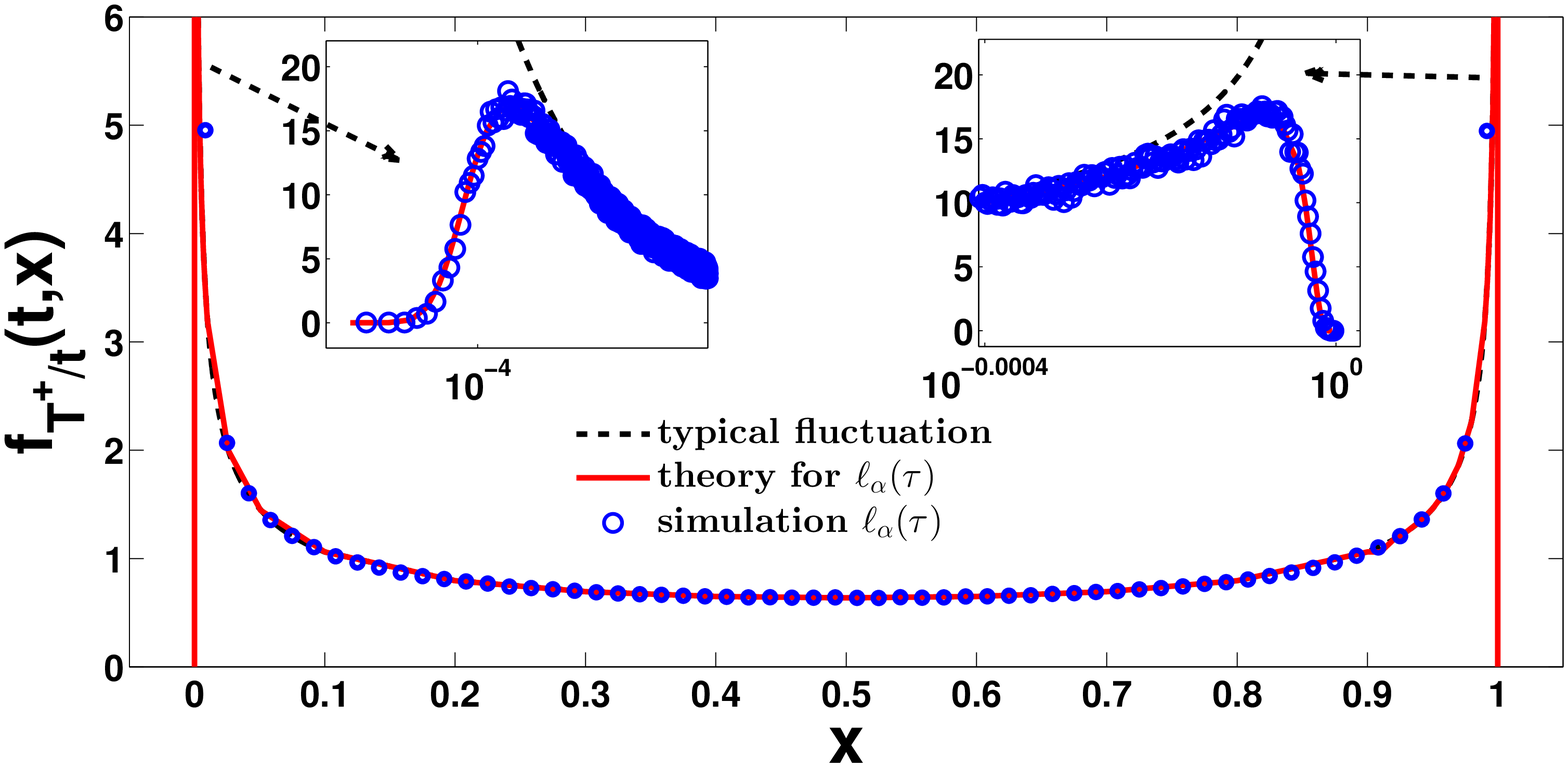}
    \caption{(color online) The occupation time density $f_{T^+/t}(t,x)$ of a renewal process  with $\phi(\tau)$ Eq.~(\ref{ldeq32011levy}). The parameters are  $\alpha=0.5$ and $t=1000$.
    The simulations, plotted by the  symbols,
   are generated by averaging $10^7$ trajectories and the curve is the  theoretical result  obtained from Eq.~(\ref{ldd107}) and symmetry of $f_{T^+}(t,x)$. Note that the results for typical fluctuations Eq.~\eqref{ldd104}, diverge on $x=0$ and $x=1$, while the large deviations theory predicts correctly finite value of the PDF.
   }\label{occpationScalelTplus}
\end{center}
\end{minipage}
\end{figure}

\subsection{Occupation time with $0<\alpha<1$}
We first consider the typical fluctuations, i.e., $T^+$ is  of the order of the measurement time $t$. Substituting  Eq.~(\ref{ldeq3201}) into Eq.~(\ref{ldd102}),
and then taking the inverse double Laplace transform, yields the PDF of $T^+/t$ \cite{Godreche2001Statistics,Lamperti1958occupation}
\begin{equation}\label{ldd103}
\begin{split}
\lim_{t\rightarrow \infty}&f_{T^+/t}(x)  \sim\frac{\sin(\pi\alpha)}{\pi} \\
    & \times \frac{x^{\alpha-1}(1-x)^{\alpha-1}}{x^{2\alpha}+(1-x)^{2\alpha}+2\cos(\pi\alpha)x^\alpha(1-x)^\alpha}
\end{split}
\end{equation}
with $0<x<1$.
It implies  that the probability distribution of the random variable $x=T^+/t$ will converge in the limit of long  $t$, to a limiting distribution which is $t$ independent.
In the particular case $\alpha=1/2$, Eq.~(\ref{ldd103})  reduces to the arcsine law on $[0,1]$
\begin{equation}\label{ldd104}
 \lim_{t\rightarrow \infty} f_{T^{+}/t}(x)\sim\frac{1}{\pi\sqrt{x(1-x)}}
\end{equation}
or
\begin{equation}\label{ldd104af}
f_{T^+}(t,T^+)=\frac{1}{\pi\sqrt{T^+(t-T^+)}}.
\end{equation}
Eq.~(\ref{ldd103}) was originally derived by Lamperti \cite{Lamperti1958occupation}; {\reform see also Darling-Kac law \cite{Darling1957On}}.
The  typical fluctuations described  by  Eq.~(\ref{ldd104}) is plotted by the  dashed (black) lines; see Figs.~\ref{occpationSmallTplus} and \ref{occpationScalelTplus}.
Besides, for $\alpha=0.5$  the typical result Eq.~(\ref{ldd104}) implies that $f_{T^+/t}(x)$ blows up when $x\rightarrow 0$ and $x\rightarrow 1$.

 Next we analyse the case of $T^{+}\ll t$, i.e., $s\ll u$. Based on  Eq.~(\ref{ldd102}), we find the infinite density
\begin{equation}\label{ldd10602}
  f_{T^+}(t,T^+)\sim \mathcal{L}^{-1}_{T^+}\left[-\frac{1}{2}+\frac{1}{1-\widehat{\phi}(u)}\right]p_0(t),
\end{equation}
where $p_0(t)$ is the survival probability defined by Eq.~(\ref{spt}). Note that Eq.~(\ref{ldd10602}) is not normalised, which is not a problem since it is valid
for $T^{+} \ll t$.

We now investigate the infinite density Eq.~\eqref{ldd10602} with two choices of $\phi(\tau)$. Similar to our previous examples the infinite density depends on the spectfics of $\phi(\tau)$ unlike the Lamperti law Eq.~(\ref{ldd103}).
Using the example of a Mittag-Leffler PDF $\phi(\tau)$, pluging Eq.~(\ref{ldbeq102hi}) into Eq.~(\ref{ldd10602}) and then taking the inverse Laplace transform
\begin{equation}\label{ldd10603}
  f_{T^+}(t,T^+)/p_0(t)\sim \frac{1}{2}\delta(T^+)+\frac{1}{\Gamma(\alpha)}(T^+)^{\alpha-1};
\end{equation}
see Fig.~\ref{occpationSmallTplus}.
The first term on the right hand side is a delta function, it describes events where the process starts at state minus and remains there for time $t$ (the factor $1/2$ is due to the initial condition, the probability of $1/2$ to start in the state up or down).
Furthermore, it is interesting to find that the typical result Eq.~(\ref{ldd103}) is  consistent with the theoretical result with Mittag-Leffler time statistics for all kinds of
$0< T^+\ll t$, not including the delta function in Eq.~\eqref{ldd10603}.

Comparing Eq.~(\ref{ldd10603}) with typical fluctuations Eq.~(\ref{ldd104af}),
we observe that for $\alpha=1/2$ the occupation time with Mittag-Leffler waiting time produces large deviations statistics  that are very similar to typical events statistics. But even in this as close as can get scenario, we find an  feature being exclusively revealed by the large deviations analysis. Namely, there is a discrete probability to find the occupation time being trapped in an initial state.
Now we derive a formal solution for the rare events.
Using the relation $1/(1-\widehat{\phi}(u))=\sum_{n=0}^{\infty}\widehat{\phi}^n(u)$ and performing the double inverse Laplace transform, leading to
\begin{equation}\label{ldd106a}
   f_{T^+}(t,T^+) \sim \Big(\frac{1}{2}\delta(T^+)+\sum_{n=1}^{\infty} \mathcal{L}_{T^+}^{-1}[\widehat{\phi}^n(u)]\Big) p_0(t).
\end{equation}
When $\phi(\tau)$ is one sided L\'{e}vy stable distribution, Eq.~(\ref{ldd106a}) reduces to
\begin{equation}\label{ldd107}
\begin{split}
   f_{T^+}(t,T^+) & \sim \Big(\frac{1}{2}\delta(T^+)+\sum_{n=1}^{\infty}\Big(\frac{1}{n^{1/\alpha}}\ell_\alpha\Big(\frac{T^+}{n^{1/\alpha}}\Big)\Big)\Big)p_0(t).
\end{split}
\end{equation}
It can be noticed that the behavior of $f_{T^+}(t,T^+)$ is determined by the shape of $\phi(\tau)$ for small $T^{+}$.  Eq.~\eqref{ldd107}, or more precisely the limit $t\to\infty$ of $f_{T^+}(t,T^+)/p_0(t)$, is the infinite density describing the occupation time statistics when $\phi(\tau)$ is the one sided L{\'e}vy distribution, see Fig.~\ref{occpationSmallTplus} for illustration.

Now we investigate  the total probability to find $0<T^{+}<T^+_1$, defined by  $P(T^{+}_1)=\int_0^{T^+_1}f _{T^+}(t,T^+)dT^+$. To simplify the discussion, we just consider Mittag-Leffler time  statistics. Using Eq.~(\ref{ldd10603}) and the asymptotic behaviors of $t^{\alpha-1}E_{\alpha,\alpha}(-t^\alpha)$ yields
\begin{equation}\label{ldocck107}
  P(T^{+}_1)\sim \frac{1}{2}p_0(t)+ \frac{\sin(\pi\alpha)}{\pi\alpha}\frac{(T^{+}_1)^\alpha}{t^\alpha}
\end{equation}
with $T^{+}_{1}\ll t$. On the other hand,
in the particular case $\alpha=1/2$,  utilizing the typical fluctuations Eq.~(\ref{ldd104af}) gives the  arcsine distribution
\begin{equation}\label{ldocck108}
  P(T^{+}_1)\sim \frac{2}{\pi}\arcsin\left({\frac{T^+_1}{\sqrt{t T^+_1}}}\right)
  %
  %
\end{equation}
It can be noted that Eq.~(\ref{ldocck108}) reduces to $2\sqrt{T^{+}_{1}}/(\pi\sqrt{t})$ for $T^{+}_1\ll t$. In this case,
we  see that Eqs.~\eqref{ldocck107}  and \eqref{ldocck108} are consistent with each other except for the the first term of Eq.~(\ref{ldocck107}).  It implies that, though Eq.~(\ref{ldd10602}) is not normaized, we can use it to calculate some observables.

\subsection{Occupation time with $1<\alpha<2$}
Now we  study the random variable $\epsilon=T^+-t/2$, shifting the symmetry axis of $f_{T^+}(t,T^+)$ to zero.
Similar to the derivation of Eq.~(\ref{ldbeqf1051}), the double Laplace transform of $f_{\epsilon}(t,\epsilon)$ is
\begin{equation}\label{lgocc601}
\begin{split}
 \widehat{f}_{\epsilon}(s,u_\epsilon) & =\Big(s+\frac{u_\epsilon(\widehat{\phi}(s+\frac{u_\epsilon}{2})-\widehat{\phi}(s-\frac{u_\epsilon}{2}))}{1-\widehat{\phi}(s+\frac{u_\epsilon}{2})\widehat{\phi}(s-\frac{u_\epsilon}{2})}\Big) \\
    &~~~\times\frac{1}{(s-\frac{u_\epsilon}{2})(s+\frac{u_\epsilon}{2})}.
\end{split}
\end{equation}
Since the sign of $\epsilon$ is not fixed, i.e., it can be positive or negative, we replace $u_\epsilon$ with $-ik$ and move to the Fourier space.
For the typical case, i.e., $ |\epsilon|\sim t^{1/\alpha}$, we find
\begin{equation}\label{lgocc602}
\widehat{f}_{\epsilon}(s,k)\sim \frac{1}{s-\frac{b_\alpha}{2\langle\tau\rangle}\left((s+\frac{ik}{2})^\alpha+(s-\frac{ik}{2})^\alpha\right)}.
\end{equation}
Taking inverse Laplace and Fourier  transform yields \cite{Schulz2015Fluctuations}
\begin{equation}\label{lgocc603}
f_{\epsilon}(t,\epsilon)\sim \left\{
                          \begin{split}
                            &\frac{C_{occ}}{t^{1/\alpha}}L_\alpha\Big(\frac{C_{occ}}{t^{1/\alpha}}|\epsilon|\Big), & \hbox{for $-t/2<\epsilon<t/2$;} \\
                           &0, & \hbox{otherwise,}
                          \end{split}
                        \right.
\end{equation}
where $C_{occ}=\langle\tau\rangle^{1/\alpha}/(2(b_\alpha|\cos(\pi\alpha/2)|)^{1/\alpha})$ and $L_\alpha(x)$ denotes the symmetric stable L\'{e}vy Law with the index of $\alpha$, so the Fourier transform of $L_\alpha(x)$ is $\exp(-|k|^\alpha)$, which is a special case of $L_{\alpha,\beta}(x)$; see Appendix \ref{ldaphy11}.
%
%

Since $0<T^+<t$, we  find that $-t/2<\epsilon<t/2$. It means that the order of $\epsilon$ can be as large as the observation time $t$. Hence to investigate the rare events, we consider $s$ is  of the order of $|k|$. By inverting the Fourier and Laplace transform, we find (see also \cite{Schulz2015Fluctuations})
\begin{equation}\label{lgocc603s}
  f_{\epsilon}(t,\epsilon)\sim \frac{\alpha b_\alpha}{t^\alpha |\Gamma(1-\alpha)|}\times\left\{
                                                                                          \begin{split}
                                                                                             &\chi\left(\frac{2|\epsilon|}{t}\right), & \hbox{$-\frac{t}{2}<\epsilon<\frac{t}{2}$;} \\
                                                                                            &0, & \hbox{otherwise}
                                                                                          \end{split}
                                                                                        \right.
\end{equation}
with $$\chi(z)=\theta(0<z\leq 1)z^{-1-\alpha}\Big(1-\frac{\alpha-1}{\alpha}z\Big).$$
We  see that $t^\alpha f_{\epsilon}(t,\epsilon)$ does not depend on the exact shape of $\phi(\tau)$ besides the parameters $\alpha$ and $b_\alpha$. Further the integral of Eq.~(\ref{lgocc603s}) with respect to $\epsilon$, in the range $\epsilon\rightarrow 0$, is divergent. Thus, $f_\epsilon(t,\epsilon)$ is a non-normalized solution since  its behavior, at $T^+\rightarrow t/2$, is non-integrable. See the discussions and numerical examples in Ref.~\cite{Schulz2015Fluctuations}.

%
\section{Discussion}
It is well known that when the averaged time interval between renewal
events diverges, i.e., $0<\alpha<1$, the typical scale of the process
is the measurement time and so observables of interest scale with $t$.
Hence  the rare fluctuations, and the far tails of the distributions
of observables considered in this paper,  have corrections when the
observable
 is of the order
of  $t^{0}$. This leads to non-normalized states which describe these
rare events. The opposite takes place when $1<\alpha<2$ namely when the mean
sojourn time is finite but the variance is diverging. Here, we have a
finite scale, but when observables like $B$,~$F$,~$Z$
or $T^{+}$ become large, namely when they are of the order of $t$,
 one naturally finds deviations from typical laws.
Since the approximation in the far tail of the distribution
 must match the typical fluctuations which are  described by fat tailed
densities,
we get by extension  non-normalized states.

{\reform The uniform approximation provided
in the text (for example  Eq.~\eqref{ldbeq304} and the corresponding Fig.~\ref{backwardScale1.5}) bridges between the typical and  rare fluctuations.  
It is obtained by matching the far tail distribution with its bulk fluctuations. Technically we find unifrom  approximation by using exact
theoretical results (see Eq.~\ref{ldf103bmlf}), an approximation where we take $s\to 0$ (meaning $t\to\infty$) leaving the second variable $u$  (corresponding for example to F) finite (see example Eq.~\eqref{ldf106}),
and for special choice of the waiting time PDF we can get the solution in terms of infinite sums (for instance Eq.~\eqref{ldti103a})}.
In principle
the uniform approximation
 can be used to calculate quantifiers  of the process like moments.
However, it is much simpler to classify observables based on their integrability with respect to the non-normalized state, as is done
in infinite ergodic theory. In the case of integrable
observables, we may use the non-normalized state for the calculation
of integrable  expectations,  some
what similar to the calculation  averages observables from normalized densities.

Importantly,  the non-normalized states are not only a tool with which we
obtain moments. As we have demonstrated both theoretically and numerically, they
describe the perfectly normalized probability density of the observables,
when the latter  are properly scaled with time (see Fig.~\ref{ForwardsmallF}).
Maybe the main achievement of this paper is that we
have obtained explicit solutions describing the rare events and this we did
with relatively simple tools.
It is rewarding that while the rare fluctuations are non-universal,
in the sense that they depend on the details of the waiting time PDF,
they can be obtained rather generally. Further, as we have shown for
the backward and forward recurrence time, the density describing the
typical fluctuations for $1<\alpha<2$
Eq.~\eqref{ldf104h02}, describes the non-typical
events for $0<\alpha<1$, all we need to do is replace the finite mean waiting time with an effective time dependent one; see  Eq.~\eqref{ldf103}.

 As mentioned in the introduction the distribution of the occupation
time for Brownian motion and random walks is the arcsine law, and
the same holds for the backward recurrence time (here $\alpha=1/2$).
While analyzing the rare events of these well known results, we see that
the large deviations for these two observables behave differently
(compare Fig.~\ref{backwardArcsin} with Fig.~\ref{occpationScalelTplus}).
For the backward recurrence time $B$ we have deviations from arcsine law
which differ for the case  $B\propto t^0$ and
 $B \propto t$ (see Fig.~\ref{backwardArcsin}). The same symmetry
breaking is not found for the occupation time since by construction
of the model the probability to be in the up (+) and down (-) state is the same
 (Fig.~\ref{occpationScalelTplus}). It should be noted
that we have worked all along with a non-equilibrium process, in the sense
that the process started at time $t=0$. Further,  for occupations time
we assumed that initially we are either in the up or down states with equal probability. In an ongoing process,
when the observation starts long after the start of the process, our results
for the rare  and even for the typical fluctuations must be modified. For example
if we start the process at time $-t_a$ before the process is observed at time
$t=0$ we expect aging effects when $0<\alpha<1$. The effect
of initial preparation on the rare fluctuations is left for
a future work, as well as the connection of results presented
here with the big jump principle \cite{Alessandro2018Single}.

\section*{Acknowledgments}
The support of Israel Science Foundation's grant 1898/17 is acknowledged, and the work was partially supported by the National Natural Science Foundation of China under
Grant No. 11671182, and the Fundamental Research Funds for the Central Universities
under Grant  No. lzujbky-2018-ot03  and lzujbky-2017-it57. W.W.
is sustained by the China Scholarship Council (CSC).

\appendix
\begin{appendices}

\section{Generation of Random Variables
}\label{generatedpower}
When generating the random variables with the PDF $\ell_\alpha(\xi)$ or $\xi^{\alpha-1}E_{\alpha,\alpha}(-\xi^\alpha)$ needed to simulate the renewal process, the Monte Carlo
statistical methods \cite{Robert2004Monte} are used.
Chambers et al. \cite{Chambers1976Methods} showed how to obtain a random variable drawn from the stable L\'{e}vy distribution with $0<\alpha<1$. 
Furthermore, Kozubowski constructed the following structural representation of a $\phi(\xi)=\xi^{\alpha-1}E_{\alpha,\alpha}(-\xi^\alpha)$ distributed random
variable $\xi$ as \cite{Kozubowski2001Fractional}
\begin{equation*}
  \xi=\sigma \eta^{1/\alpha},
\end{equation*}
where $\sigma $  is a random number from the exponential distribution with mean parameter $1$, and $ \eta$ has the PDF
\begin{equation*}
  f(\eta)=\frac{\sin(\pi\alpha)}{\alpha\pi(\eta^2+2\eta\cos(\pi\alpha)+1)}
\end{equation*}
with $0<\alpha<1$ and $\eta>0$.

\section{The derivation of some important formulas}\label{ldbasic}
Now we give a brief  account on some main equations used in the paper. Let us consider a process starting at time $t=0$ and define $Q_N(t)$, which is  the probability that the $N$-th event happens at time $t=t_N=\tau_1+\tau_2+, \ldots, +\tau_N$ (see Fig.~\ref{rennew}). Note that an important relation between $Q_N(t)$ and $Q_{N-1}(t)$ is Eq.~\eqref{ldnu10111aop}.
%
%
Then using the convolution theorem of Laplace transform and $Q_1(\tau)=\phi(\tau)$,
it follows that
\begin{equation}\label{ldbasic102a}
\widehat{Q}_N(s)=\widehat{\phi}^N(s).
\end{equation}
Another important equality is the conditional probability density of the forward recurrence time $F$ given that exactly $N$ events occurred before time $t$, defined by
\begin{equation}\label{ldbasic102}
f_N(t,F)=\int_0^tQ_N(\tau)\phi(t-\tau+F)d\tau.
\end{equation}
Note that the forward recurrence time is given by $f_{F}(t,F)=\sum_{N=0}^{\infty}f_N(t,F)$, i.e., Eq.~(\ref{ldf100}). The Laplace transform of $f_F(t,F)$ with respect to $t$ follows from the shift theorem of Laplace transform and reads
\begin{equation*}
  \widehat{f}_F(s,F)=\frac{1}{1-\widehat{\phi}(s)}\exp(sF)\int_F^\infty\phi(z)\exp(-sz)dz.
\end{equation*}
Then, taking Laplace transform and using partial integration, lead to the final result Eq.~(\ref{ldf101}).

Now, our aim  is to obtain  the PDF of the occupation time. Let $Q_N(t,T^+)$ be the PDF of the occupation time just arriving at $T^+$ at time $t$ after finishing $N$ steps. $Q_N^{\pm}(t,T^+)$ is
\begin{equation}\label{ldbasic103}
\begin{split}
  Q_{N+2}^{+}(t,T^+) & =\int_0^{T^+}\int_0^tQ_N(t-\tau-z,T^+-z) \\
    & \times\phi(z)\phi(\tau)d\tau dz+q\delta(t)\delta(t-T^+)
\end{split}
\end{equation}
and
\begin{equation}\label{ldbasic103a}
\begin{split}
  Q_{N+2}^{-}(t,T^+) & =\int_0^{T^+}\int_0^tQ_N(t-\tau-z,T^+-z) \\
    & \times\phi(z)\phi(\tau)d\tau dz+(1-q)\delta(t)\delta(T^+),
\end{split}
\end{equation}
where $\pm$ in the superscript of $Q_{N+2}^{\pm}(t,T^+)$ means that the initial state of the particle  is $\pm$. $q$ is the probability that the initial state is  $+$, with $0\leq q\leq1$.  In double Laplace space, representation of the above  Eq.~(\ref{ldbasic103}) takes an especially simple form
\begin{equation}\label{ldbasic104}
\widehat{Q}_{N+2}^{+}(s,u)=\frac{q}{1-\widehat{\phi}(s+u)\widehat{\phi}(s)}.
\end{equation}
Then the PDF  $f_{T^+}(t,T^+)$ is
\begin{equation*}
\begin{split}
  &f_{T^+}(t,T^+)  =\sum_{N=0}^\infty \Big(\\
  &\int_0^t\int_0^{T^+}Q_N^+(t-\tau-z,T^+-z)\int_z^{\infty}\phi(y)dydzdt \\
    & +\int_0^t\int_0^{T^+}Q_N^+(t-\tau-z,T^+-z)\phi(z)\int_\tau^{\infty}\phi(y)dydzdt\\
    &+\int_0^t\int_0^{T^+}Q_N^-(t-\tau-z,T^+-z)\phi(\tau)\int_z^{\infty}\phi(y)dydzdt\\
    &+\int_0^tQ_N^-(t-\tau,T^+)\int_\tau^{\infty}\phi(y)dydt\Big).
\end{split}
\end{equation*}
Taking  double Laplace transform, summing the infinite terms, and  then from Eq.~(\ref{ldbasic104}) it follows that
\begin{equation}\label{ldbasic105}
\begin{split}
  \widehat{f}_{T^+}(s,u) &=\frac{u(1-q-q\widehat{\phi}(s+u))(1-\widehat{\phi}(s))}{s(s+u)(1-\widehat{\phi}(s)\widehat{\phi}(s+u))} \\
    &~~~~~~+\frac{s}{s(s+u)}.
\end{split}
\end{equation}
For $q=1/2$, Eq.~(\ref{ldbasic105}) reduces to Eq.~(\ref{ldd102}) and the corresponding typical fluctuations are studied in Refs. \cite{Godreche2001Statistics,Gennady2005Power}.

%
%
%
%
%

\section{Some properties of Stable distribution}\label{ldaphy11}
Now we discuss the series representation and the asymptotic behavior of stable distribution $L_{\alpha,\beta}(x)$ \cite{Schneider1986Stable,Metzler2000random,Metzler2004The}. The corresponding PDF $L_{\alpha,\beta}(x)$ is given by the inverse Fourier transform
\begin{equation}\label{lglevydis}
L_{\alpha,\beta}(x)=\frac{1}{2\pi}\int_0^\infty \exp\Big(-ikx-c|k|^\alpha\Big(1+i\beta\frac{z}{|z|}h(z,\alpha)\Big)\Big)dk,
\end{equation}
where $\alpha,\beta,c$ are constants and
\begin{equation*}
  h(z,\alpha)=\left\{
                \begin{split}
                  &\tan\Big(\frac{\pi\alpha}{2}\Big), & \hbox{$\alpha\neq 1$;} \\
                  &\frac{\pi}{2}\log(z), & \hbox{$\alpha=1$.}
                \end{split}
              \right.
\end{equation*}
Especially, for $\beta=0$ and $c=1$, Eq.~(\ref{lglevydis}) reduces to the symmetric stable distribution $L_{\alpha}(x)$. For simplification of analysis,
let $\beta=1$, $\alpha\neq1$, and $c=-\cos(\pi\alpha/2)$,
\begin{equation*}
\begin{split}
 L_{\alpha,1}(x) =\frac{1}{2\pi}\int_{-\infty}^{\infty}\exp(-ikx)\exp[(ik)^\alpha]dk.
\end{split}
\end{equation*}
Expanding the integrand in the right hand side as a Taylor series in $x$ yields the convergent series
\begin{equation}\label{ldappell102}
\begin{split}
L_{\alpha,1}(x)&= \sum_{n=0}^{\infty}\frac{(-1)^n}{\alpha \pi (2n)! }\Gamma\left(\frac{2n+1}{\alpha}\right)\cos(g(n,\alpha))x^{2n} \\
    &- \sum_{n=0}^{\infty}\frac{\sign\!(x)(-1)^n\Gamma(\frac{2n+2}{\alpha})}{\pi\alpha(2n+1)!}
    \\
    &
    \times\sin\Big(g\Big(n+\frac{1}{2},\alpha\Big)\Big)x^{2n+1},
\end{split}
\end{equation}
where $g(n,\alpha)=(2n+1)\pi(1/2-1/\alpha)$ and $\sign\!(x)=x/|x|$ for $|x|>0$ and zero otherwise.  Especially, for $x\rightarrow 0$, Eq.~(\ref{ldappell102}) reduces to
\begin{equation*}
  L_{\alpha,1}(x)\sim \frac{\Gamma(\frac{1}{\alpha})}{\alpha\pi}\cos(g(0,\alpha)),
\end{equation*}
which is a constant and strictly less than  $L_{\alpha}(0)$ for $\alpha<2$.
Furthermore,
the asymptotic behavior of $L_{\alpha,1}(x)$ is
\begin{equation}\label{ldappell103}
   L_{\alpha,1}(x) \sim \sum_{n=1}^{\infty}\frac{\Gamma(1+\alpha n)\sin(\alpha n\pi)}{2\pi n!|x|^{1+\alpha n}}(-1+\sign\!(x)),
\end{equation}
 it implies that $L_{\alpha,1}(x)\sim |x|^{-1-\alpha}/\Gamma(-\alpha)$, being the same as the left hand side of the tail of the symmetric  L\'{e}vy stable distribution, for $x\rightarrow -\infty$ and the tails of $L_{\alpha,1}(x)$ are asymmetric with respect to $x$.


\section{The calculation of $q$ order moments for forward  recurrence time with $\alpha>1$}\label{ldappfor1}
Using the calculated results of $f_F(t,F)$, we  study  fractional moments $\langle F^q \rangle$.
First, we obtain the low order moments with $\alpha>1$, i.e., $q<\alpha-1$. Using Eqs.~(\ref{ldf104h02}) and (\ref{ldf108}), and utilizing integration by parts
\begin{equation}\label{ldap101}
 \langle F^q\rangle\sim \frac{1}{(q+1)\langle\tau\rangle}\int_0^{\infty}F^{q+1}\phi(F)dF.
\end{equation}
We  notice that the right hand side of Eq.~(\ref{ldap101}) is a finite number due to $q-\alpha<-1$. Then, we  discuss the case of $q>\alpha-1$. According to  Eq.~(\ref{ldf107f})
\begin{equation}\label{ldap1022}
\begin{split}
  \langle F^q\rangle & \sim\lim_{z\rightarrow 0}\int^{\infty}_{z} (F^{-\alpha}-(F+t)^{-\alpha})\frac{b_\alpha F^q}{|\Gamma(1-\alpha)|\langle\tau\rangle} dF\\
    & \sim \lim_{z\rightarrow 0} \frac{b_\alpha z^{1-\alpha+q}}{|\Gamma(1-\alpha)|\langle\tau\rangle(1+q-\alpha)}\\
    &~~~\times F\Big(\alpha,-1+\alpha-q,\alpha-q; -\frac{t}{z}\Big),
\end{split}
\end{equation}
where $F(\alpha, \beta, \gamma; x)$ is the hypergeometric function \cite{Abramowitz1984Handbook,Seaborn1991Hypergeometric}, defined by
\begin{equation}\label{ldaphy101}
F(\alpha, \beta, \gamma; x)=1+\sum_{n=1}^{\infty}\frac{(\alpha)_n (\beta)_n}{(\gamma)_n}\frac{x^n}{n!}
\end{equation}
with $(\alpha)_n=\alpha(\alpha+1)\ldots(\alpha+k-1)$.
Note that the asymptotic behavior of $F(\alpha, \beta, \gamma; -x)$ is
\begin{equation}\label{ldap10223}
 F(\alpha, \beta, \gamma; -x)\sim x^{-\alpha}\frac{\Gamma(\beta-\alpha)\Gamma(\gamma)}{\Gamma(\beta)\Gamma(\gamma-\alpha)}+x^{-\beta}\frac{\Gamma(\alpha-\beta)\Gamma(\gamma)}{\Gamma(\alpha)\Gamma(\gamma-\beta)}
\end{equation}
with $x>0$.
Using Eq.~(\ref{ldap10223}), the dominant term of Eq.~(\ref{ldap1022}) gives
\begin{equation}\label{ldap10223e}
 \langle F^q\rangle\sim \frac{b_\alpha\Gamma(1+q)\Gamma(\alpha-q)}{|\Gamma(1-\alpha)|\langle\tau\rangle(1+q-\alpha)\Gamma(\alpha)}t^{1-\alpha+q},
\end{equation}
keep in mind that substituting uniform approximation Eq.~(\ref{ldf106}) into Eq.~(\ref{ldf108}) yields the same result as Eq.~(\ref{ldap10223e}).
In the particular case $q=1$, we have $\langle F\rangle\sim t^{2-\alpha}$.
Besides, when $q>\alpha$, as expected $\langle F^q\rangle$ diverges.

%
%
%
%
%
\section{$q$ order moments with $\alpha<1$ }\label{qmentalpha0}
We now  study  the fractional moments of $B$. Note that  $ B^q$ with $q>\alpha$, are non-integrable with respect to the non-normalized density Eq.~(\ref{ldbeq103}). We find that
the fractional moments of $B$  are governed by the typical fluctuations Eq.~(\ref{ldbeq105a}), namely
\begin{equation}\label{alpha0momentsback}
\begin{split}
 \langle B^q\rangle&=\int_0^\infty B^qf_B(B,t)dB \\
    & \sim\frac{\sin(\pi\alpha)\Gamma(\alpha)\Gamma(1-\alpha+q)}{\pi\Gamma(1+q)}t^q.
\end{split}
\end{equation}
We  check this result in the following:
for a natural number $q$,  expanding Eq.~(\ref{ldbeq101}) as a Taylor series in $u$, and performing the inverse Laplace transform term by term, we  obtain the corresponding  moments, which are the same as Eq.~(\ref{alpha0momentsback}).

Let us consider another interesting observable, i.e., the  moments of $N$. Using Eqs.~(\ref{ldnu106}) and (\ref{ldf108})
\begin{equation}\label{fracmomentsN0}
\begin{split}
 \langle N^q\rangle&\sim\int_0^\infty \frac{tN^q}{\alpha N^{1+1/\alpha}b_\alpha^{1/\alpha}}\ell_\alpha\Big(\frac{t}{(Nb_\alpha)^{1/\alpha}}
 \Big)dN \\
    & =\frac{\int_0^\infty \xi^{-\alpha q}\ell_\alpha(\xi)d\xi}{(b_\alpha)^q}t^{\alpha q}.
\end{split}
\end{equation}
In the particular case $q\to 0$, the normalized condition is found, namely, $\langle N^0\rangle=1$.

To summarize, if $\alpha<1$, then for all observables in this paper, i.e., $N, F, B, Z$, and $T^+$, the moments (if they exist) are obtained by the PDF describing the typical fluctuations. Note that for $F$ and $ Z$, high order ($q>\alpha$) moments diverge.
One may wonder in what sense is Eq.~(\ref{ldbeq103}) an infinite density? For that we consider the observable $\Theta(B_1<B<B_2)$ with $B_1, B_2\ll t$, where $\Theta(B_1<B<B_2)$ is one if the condition holds. Then
\begin{equation*}
  \begin{split}
    \langle \Theta(B_1<B<B_2)\rangle & =\int_{0}^{\infty}\Theta(B_1<B<B_2)f_B(t,B)dB\\
      &\sim\frac{1}{\langle\tau^*\rangle}\int_{B_1}^{B_2}\int_B^{\infty}\phi(y)dydB,
  \end{split}
\end{equation*}
where $\langle\tau^*\rangle$, defined below Eq.~(\ref{ldf103}), is the effective average waiting time. In other words, the observable $\Theta(B_1<B<\theta_2)$ is integrable with respect to the non-normalized density, and hence the latter is used for the calculation of the average $\Theta(B_1<B<\theta_2)$.

\end{appendices}

\bibliographystyle{prestyle}

\end{document}